\documentclass[12pt,preprint]{aastex}

\newcommand{\myemail}{ana@taurus.phys.s.u-tockyo.ac.jp}

\shorttitle{SiO as a tracer of past protostellar activity}
\shortauthors{L\'opez-Sepulcre et al.}

\begin{document}


\title{The role of SiO as a tracer of past star-formation events: \\
    The case of the high-mass protocluster NGC\,2264-C}


\author{Ana L\'opez-Sepulcre\altaffilmark{1}, Yoshimasa Watanabe\altaffilmark{1}, Nami Sakai\altaffilmark{2}, Ryuta Furuya\altaffilmark{1}, Osamu Saruwatari\altaffilmark{1}}
%\affil{Department of Physics, The University of Tokyo, 7-3-1 Hongo, Bunkyo-ku, Tokyo, 113-0033, Japan}

\email{\myemail}

\and

\author{Satoshi Yamamoto\altaffilmark{1}}
%\affil{Space Telescope Science Institute, Baltimore, MD 21218}


\altaffiltext{1}{Department of Physics, The University of Tokyo, 7-3-1 Hongo, Bunkyo-ku, Tokyo, 113-0033, Japan}
\altaffiltext{2}{RIKEN, S407 Chemistry and Materials Physics Building, 2-1 Hirosawa, Wako, Saitama, 351-0198, Japan}


\begin{abstract}
NGC\,2264-C is a high-mass protocluster where several star-formation events are known to have occurred. To investigate whether past protostellar activity has left a chemical imprint in this region, we mapped it in SiO($J = 2-1$), a shock tracer, and several other molecular lines with the Nobeyama\,45\,m telescope. Our observations show the presence of a complex network of protostellar outflows. The strongest SiO emission lies beyond a radius of $\sim 0.1$\,pc with respect to the center of the clump, and is characterized by broad ($> 10$\,km\,s$^{-1}$) lines and abundances of $\sim 1.4 \times 10^{-8}$ with respect to H$_2$. Interestingly, SiO appears relatively depleted ($\chi_\mathrm{SiO} \sim 4 \times 10^{-9}$) within this radius, despite it being affected by molecular outflow activity. We attribute this to fast condensation of SiO back onto dust grains and/or rapid gas-phase destruction of SiO, favored by the high density present in this area ($> 10^6$\,cm$^{-3}$). Finally, we identify a peripheral, narrow-line ($\sim 2$\,km\,s$^{-1}$) component, where SiO has an abundance of a few times 10$^{-11}$. After considering different options, we conclude that this weak emission may be tracing protostellar shocks from the star formation episode that preceded the current one, which have decelerated over time and eventually resulted in SiO being largely depleted/destroyed. Alternatively, a population of unresolved low-mass protostars may be responsible for the narrow SiO emission. High-angular resolution observations are necessary to distinguish between these two possibilities and thus understand the role of SiO as a chemical tracer of past star-formation episodes in massive protoclusters.
\end{abstract}


\keywords{stars: formation --- ISM: abundances --- ISM: individual (NGC 2264) --- ISM: jets and outflows --- ISM: molecules}



\section{Introduction}\label{intro}

\subsection{High-mass star and cluster formation}

There is substantial evidence that most stars in our Galaxy form in clustered environments rather than in isolation (e.g. \citealp{ll03}). These populated environments, or \textit{protoclusters}, are excellent laboratories to study a variety of physical and chemical processes, such as cloud fragmentation and feedback from young (proto-)stars on their surroundings (see \citealp{krum14} for a review). Moreover, high-mass stars ($M_\star > 8$\,M$_\odot$) tend to form in dense clusters. The study of \textit{protoclusters} is therefore essential if we aim to investigate the dominant star-formation mode in our Galaxy and, in particular, the formation of high-mass stars, which still poses some major questions.

One of such major questions is how a massive molecular core can evolve without fragmenting into smaller cores. Some solutions have been proposed, such as radiative feedback or heating from a first generation of low-mass (proto-)stars (\citealp{krum07}, \citealp{krum08}, \citealp{long11}). If this is the case, an episode of low-mass star formation should precede the formation of massive stars. This implies that the molecular cloud hosting such a protocluster should resist global gravitational collapse long enough to allow the formation of more than one generation of protostars. In short, some forces must be present that counterbalance gravity on protocluster scales ($\sim\,1\,$pc) for several free-fall times.

Turbulent motions within molecular clouds may provide an effective source of energy acting against gravity. In particular, protostellar feedback has been proposed to be a potentially dominant source of turbulence replenishment in active protoclusters. Li \& Nakamura~\cite{ln06} performed 3D magneto-hydrodynamic simulations of cluster formation, and found that the initial cloud turbulence is quickly replaced by motions generated by outflows, thus keeping the cloud stable against gravitational collapse long after the initial turbulence has dissipated. Other theoretical and observational studies that followed lend further support to this finding (e.g. \citealp{nl07}, \citealp{sw07}, \citealp{matzner07}, \citealp{nakamura11}). In the present study, we target an active high-mass protocluster where such a scenario may be taking place: NGC\,2264-C.

\subsection{The target: NGC\,2264-C}

NGC\,2264C is a molecular clump located at a distance of 740\,pc \cite{kame14}, known to be surrounded by a large cluster of pre-main-sequence stars (e.g. \citealp{sung97}, \citealp{rebull02}, \citealp{sung08}). It contains multiple dense cores amounting to a total mass of 1650\,M$_\odot$ (Peretto et al.~\citealp{peretto06}), and very complex CO line wing components from multiple molecular outflows (Maury et al.~\citealp{maury09}). At the geometrical center of the clump lies the most massive molecular core in this site, CMM3 (50\,M$_\odot$), which is thought to be the precursor of a high-mass star. Saruwatari et al.~\cite{saru11} proposed it to be a very young massive protostellar candidate based on the discovery of a compact CO outflow with a dynamical age of a few hundred years. More recently, Watanabe et al.~\cite{wata15} reported further evidence of the core's youth based on its chemical composition, which displays high abundances of deuterated molecules and carbon chains relative to those measured in Orion\,KL, and low abundances of complex organic molecules.

In summary, there is considerable evidence suggesting that NGC\,2264C has experienced more than one episode of star formation, with one of the latest potentially being that of a massive star. This region therefore appears to be a clear case where low-mass star formation precedes high-mass star formation, and where protostellar outflows may have provided the necessary amount of turbulence to avoid fast cloud collapse.

\subsection{SiO, a tracer of protostellar activity}

Protostellar outflow activity may leave a molecular trace on its host cloud that might take a long time to erase completely. Our goal here is to assess whether SiO can be used as such a tracer in NGC\,2264C. If so, it could provide an astrochemical means to evaluate the presence of old protostellar activity and past star-formation episodes in other high-mass star-formation sites hosting very young massive cores.

The choice of SiO for this study relies on its association with shocks, in particular those originating from protostellar outflows (e.g. \citealp{gueth98}, \citealp{clau07}, \citealp{tafalla15}). Indeed, the abundance of SiO in the gas phase can be enhanced by several orders of magnitude in protostellar shocks due to sputtering of Si atoms from grains or direct vaporisation of SiO from grain mantles (e.g. \citealp{gusdorf08}, \citealp{guillet09}, \citealp{anderl13}). Since the SiO depletion time back onto the grains, as well as the time scale for its gas-phase destruction, is quite fast in protostellar regions (typically $\sim\,10^4$\,years; \citealp{bergin98}, \citealp{pdf97}), this molecule should be enhanced preferentially close to shocks and should therefore be associated with high velocities. However, a typically extended, narrow-line component has also been detected in a number of star-forming regions. This may be associated to past high-velocity shocks that have decelerated with time, and that have not yet suffered complete depletion (e.g. Codella et al.~\citealp{clau99}), although other explanations are possible, such as low-velocity shocks caused by large-scale flow collisions during global collapse or by dynamical interaction of two clouds, for instance (Jim\'enez-Serra et al.~\citealp{js10}, Sanhueza et al.~\citealp{sanhueza13}).

With all the above in mind, we have mapped NGC\,2264C in SiO($J = 2-1$) and other molecular tracers with the Nobeyama~45-m telescope. The observations are described in Sect.\,\ref{obs}. Maps and spectra are presented in Sect.\,\ref{results}, and a quantitative analysis is provided in Sect.\,\ref{derivation}. We discuss our results in Sect.\,\ref{discussion} and conclude in Sect.\,\ref{conclusions}.

\section{Observations}\label{obs}

%% In a manner similar to \objectname authors can provide links to dataset
%% hosted at participating data centers via the \dataset{} command.  The
%% second curly bracket argument is printed in the text while the first
%% parentheses argument serves as the valid data set identifier.  Large
%% lists of data set are best provided in a table (see Table 3 for an example).
%% Valid data set identifiers should be obtained from the data center that
%% is currently hosting the data.
%%
%% Note that AASTeX interprets everything between the curly braces in the 
%% macro as regular text, so any special characters, e.g. "#" or "_," must be 
%% preceded by a backslash. Otherwise, you will get a LaTeX error when you 
%% compile your manuscript.  Special characters do not 
%% need to be escaped in the optional, square-bracket argument.

We obtained maps of NGC\,2264-C in SiO($J = 2-1$), CH$_3$OH(2$_{0}-1_{0}$~A$^+$), and H$^{13}$CO$^{+}$($J = 1-0$) with the Nobeyama 45\,m telescope at the Nobeyama Radio Observatory (NRO) in 2009 March (SiO) and 2011 April - May (CH$_3$OH and H$^{13}$CO$^+$).  The side-band-separating (2SB) mixer receiver T100H/V was used as a frontend with typical system noise temperature of 160 -- 360\,K and 160 -- 270\,K for the SiO and the CH$_3$OH/H$^{13}$CO$^{+}$ observations, respectively.  The beam size of the Nobeyama 45\,m telescope was $19''$ at 86\,GHz.  The main beam efficiency ($\eta_{\rm mb}$) was 42\%.  The backends used were the accousto-optical radio spectrometers AOS-H, whose bandwidth and frequency resolution were 40\,MHz and 37\,kHz, respectively.  The position switch method was employed with an off position of (R.A. offset, Dec. offset) = ($+2'$, $+30'$). The intensity scale was calibrated to the antenna temperature ($T^{*}_{\rm A}$) scale by the chopper-wheel method and its accuracy is estimated to be 20\,\%.  The telescope pointing was checked every hour by observing the SiO maser source GX\,Mon. The pointing accuracy is confirmed to be better than $5''$. The mapped area is about $4.2' \times 4.2'$ centered at (R.A., Dec.) = (06$^\mathrm{h}$41$^\mathrm{m}$12.3$^\mathrm{s}$, +09$^\circ$29$'$12$''$) with a grid spacing of $19''$.  At the central $1.4' \times 1.6'$ of the SiO map, the grid spacing is a nyquist grid of $9.5''$.  

The data were reduced using the software package NEWSTAR, developed by NRO. After subtracting the baseline via 1st order polynomial fitting, final spectra were obtained for each position in the map. The $T^{*}_{\rm A}$ scale was converted to the main-beam brightness temperature ($T_{\rm mb}$) scale by $T_{\rm mb} = T^{*}_{\rm A}/\eta_{\rm mb}$.  

We observed CO($J = 1-0$) in 2012 March with the Nobeyama 45\,m telescope. The T100H/V receiver was used as front end with a typical system noise temperature of 250 -- 280\,K. The resulting beam size of the NRO 45\,m telescope was $15''$ at 115\,GHz. The main beam efficiency ($\eta_{\rm mb}$) was 33\,\%.  The backends used were the SAM45 auto-correlators, whose bandwidth and frequency resolution were 125\,MHz and 30.52\,kHz, respectively. The on-the-fly mapping method \cite{sawada08} was employed to cover a $3' \times 4'$ area centered at (R.A., Dec.) = (06$^\mathrm{h}$41$^\mathrm{m}$12.3$^\mathrm{s}$, +09$^\circ$29$'$12$''$), with the off position being (R.A. offset, Dec. offset) = ($+2'$, $+30'$).  In order to reduce the noise derived from scanning effects, we obtained two CO maps with two different orthogonal scanning directions and combined them in the data reduction procedure. The intensity scale was calibrated to the $T^{*}_{\rm A}$ scale by the chopper-wheel method.  The $T^{*}_{\rm A}$ scale was converted to the $T_{\rm mb}$ scale using $T_{\rm mb} = T^{*}_{\rm A}/\eta_{\rm mb}$. The pointing accuracy is better than $5''$ based on observations of the SiO maser source GX\,Mon.

The data were reduced using the software package NOSTAR \cite{sawada08}, developed by NRO. After subtracting the baseline via 1st order polynomial fitting, the data were re-sampled with a grid size of $6''$ and integrated. In the re-sampling procedure, the angular resolution was broadened to $\sim 19.5''$.

The observed molecular transitions are listed in Table~\ref{tline}, along with their properties. The average systemic velocity of NGC\,2264-C is 7.5\,km\,s$^{-1}$.

\begin{deluxetable}{llcccccc}
\tabletypesize{\footnotesize}
\tablewidth{0pt}
\tablecolumns{8}
\tablecaption{Molecular lines observed\tablenotemark{a}\label{tline}}
\tablehead{
\colhead{Molecule}  & \colhead{Transition}  & \colhead{$\nu_0$} & \colhead{$E_\mathrm{u}$}  & \colhead{$A_{ul}$} & \colhead{$\delta V$\tablenotemark{b}}  & \colhead{$\theta_\mathrm{beam}$} & \colhead{RMS noise\tablenotemark{c}}\\
\colhead{}  & \colhead{}  & \colhead{(MHz)} & \colhead{(K)}  & \colhead{(s$^{-1}$)} & \colhead{(km\,s$^{-1}$)}  & \colhead{($''$)} & \colhead{(mK)}}
\startdata
H$^{13}$CO$^{+}$ & $J = 1 - 0$ & 86754.288 & 4.2 & $3.9 \times 10^{-5}$ & 0.5 & 19.5 & 91\\
SiO & $J = 2 - 1$ & 86846.985 & 6.3 & $2.9 \times 10^{-5}$ & 0.5 & 19.5 & 54\\
CH$_3$OH & 2$_{-1} - 1_{-1}$ E & 96739.362 & 12.5 & $2.6 \times 10^{-6}$ & 0.5 & 17.5 & 50\\
CH$_3$OH & 2$_{0} - 1_{0}$ A$^+$ & 96741.375 & 7.0 & $3.4 \times 10^{-6}$ & 0.5 & 17.5 & 50\\
CH$_3$OH & 2$_{+0} - 1_{+0}$ E & 96744.550 & 20.1 & $3.1 \times 10^{-6}$ & 0.5 & 17.5 & 50\\
CO & $J =1 - 0$ & 115271.20 & 5.5 & $7.2 \times 10^{-8}$ & 0.5 & 14.7 & 530\\
\enddata
\tablenotetext{a}{The three CH$_3$OH lines are blended with each other. Throughout this paper, we use only the central velocities of the (2$_{0} - 1_{0}$)A$^+$ transition.}
\tablenotetext{b}{Channel width after spectral smoothing for a higher signal-to-noise ratio}
\tablenotetext{c}{Main beam temperature ($T_\mathrm{mb}$) units}
\end{deluxetable}

%% In this section, we use  the \subsection command to set off
%% a subsection.  \footnote is used to insert a footnote to the text.


%% This section also includes several of the displayed math environments
%% mentioned in the Author Guide.

\section{Results}\label{results}

\subsection{Integrated maps and spectra} \label{maps}

\begin{figure*}
\epsscale{0.78}
%\epsscale{1.56}
\plotone{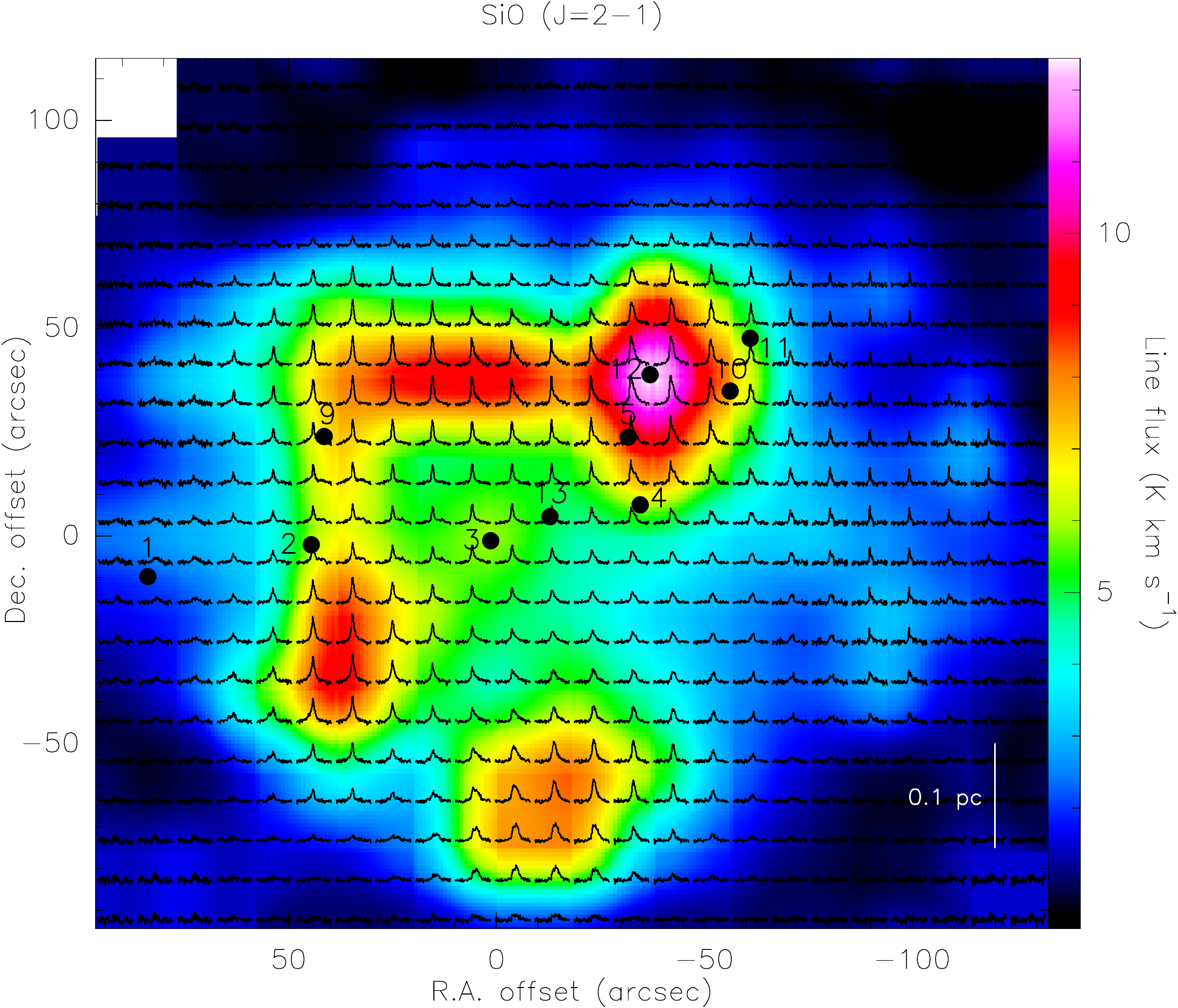}
\epsscale{0.47}
%\epsscale{0.99}
\plotone{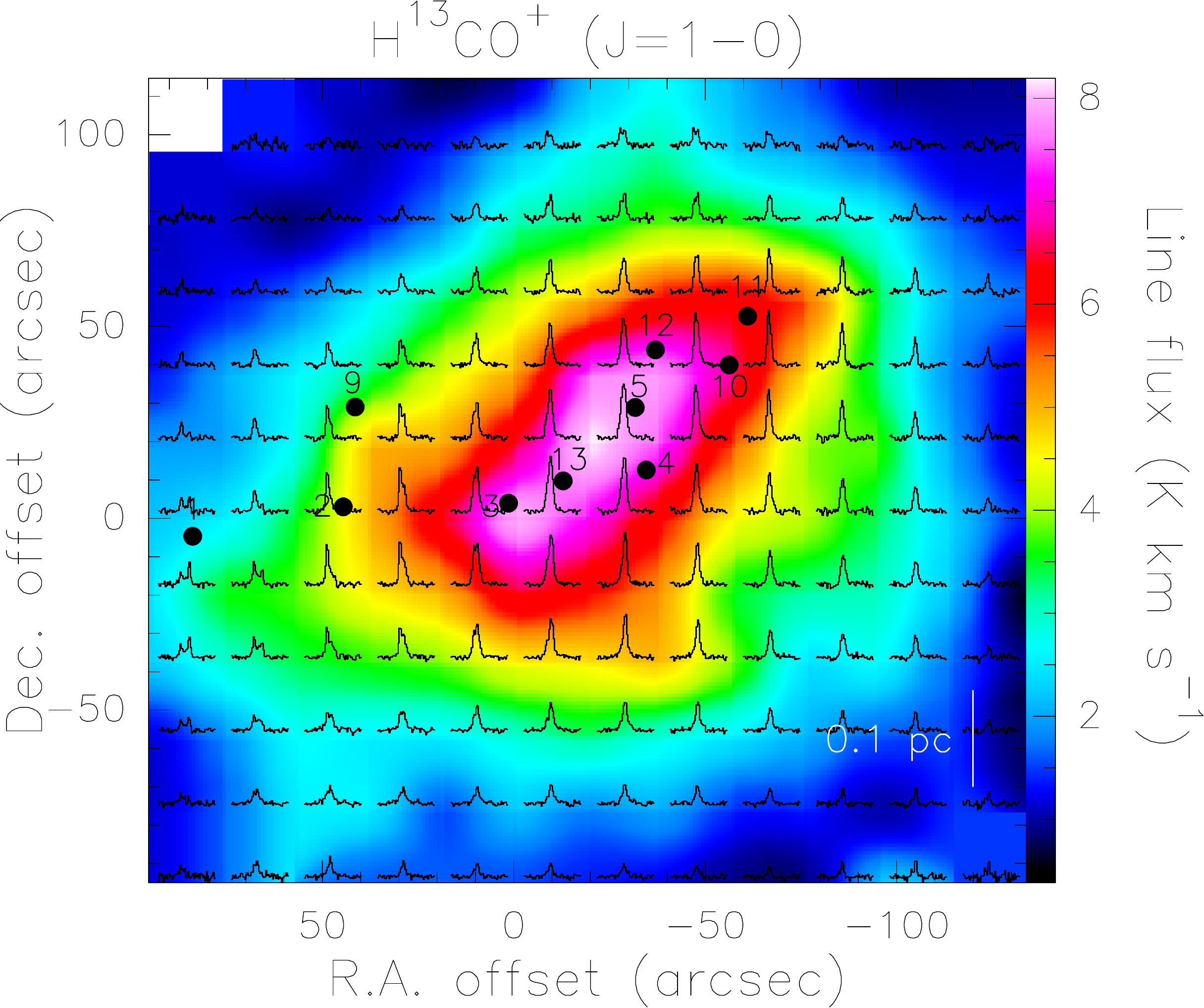}
\plotone{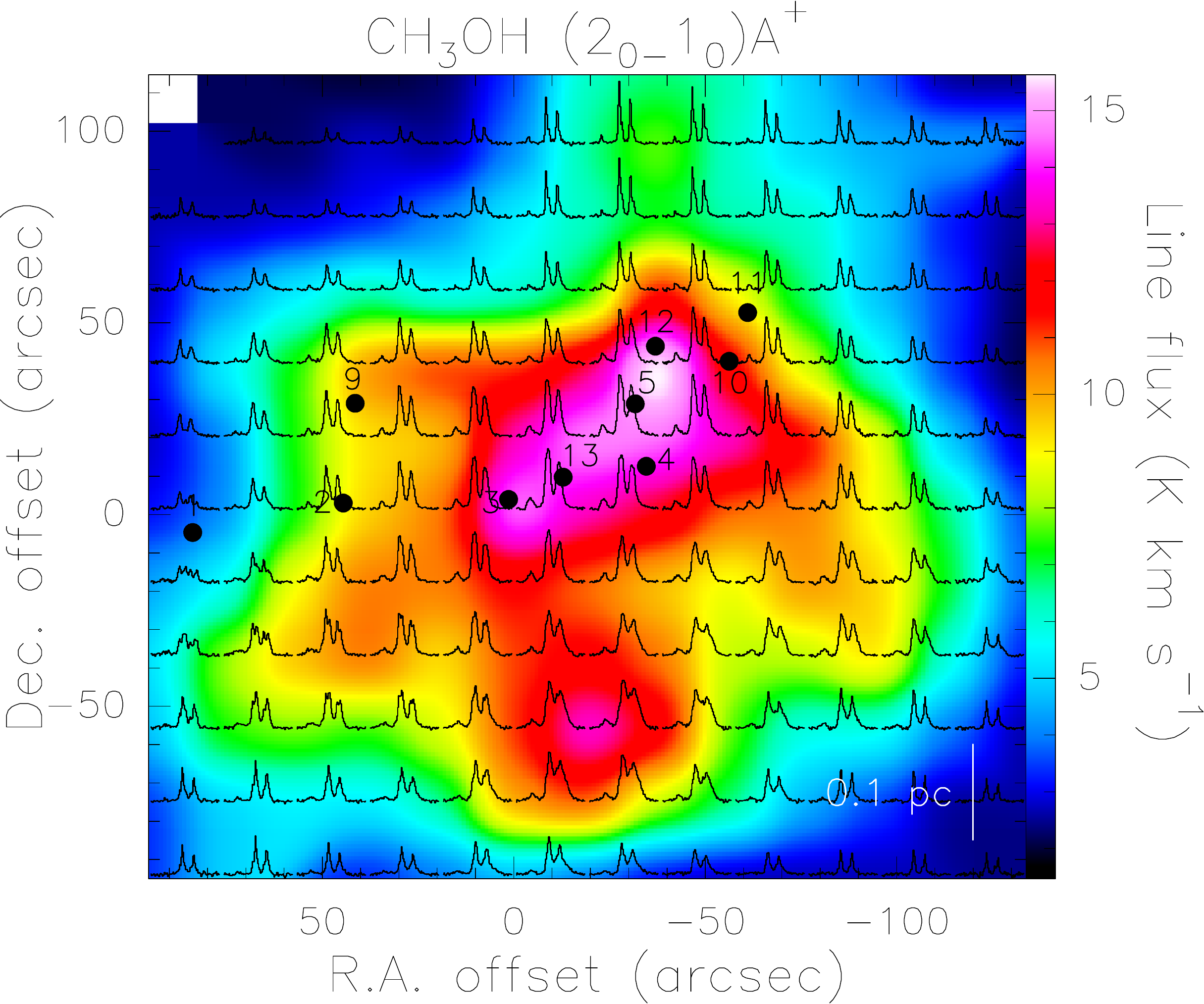}
\caption{Spectral line maps of SiO($J = 2-1$), H$^{13}$CO$^+$($J = 1-0$), and CH$_3$OH(2$_{0} - 1_{0}$ A$^+$), at intervals of 9.5$''$, 19$''$, and 19$''$, respectively, overlaid on the corresponding velocity-integrated maps (color scale). The velocity ranges used to obtain the maps are [--1.3,20.3]\,km\,s$^{-1}$ for SiO and [+5.5,+10.5]\,km\,s$^{-1}$ for H$^{13}$CO$^+$ and CH$_3$OH (see Table\,\ref{tmaps}). Black circles mark the positions of the millimeter cores identified by Peretto et al. (\citealp{peretto06}, \citealp{peretto07}). The reference (0$''$,0$''$) position corresponds to CMM3: R.A.(J2000)\,=\,06$^\mathrm{h}$41$^\mathrm{m}$12.3$^\mathrm{s}$, Dec.(J2000)\,=\,09$^\circ$29$'$11.90$''$.
\label{fsptmap1}}
\end{figure*}

Figure\,\ref{fsptmap1} presents the integrated SiO($J = 2-1$) map of NGC\,2264-C (color scale), obtained by integrating the emission of each velocity channel with a signal-to-noise (S/N) ratio above or equal to 5 in the channel map. Overlaid on the map are the corresponding spectra at each offset position observed. Analogous maps are shown in the lower panels of Fig.\,\ref{fsptmap1} for H$^{13}$CO$^+$($J = 1-0$) and CH$_3$OH(2$_{0} - 1_{0}$~A$^+$) (hereafter H$^{13}$CO$^+$ and CH$_3$OH, respectively).

Columns 2 and 3 in Table\,\ref{tmaps} list the velocity ranges used to obtain the integrated maps and the resulting RMS noise. Note the wide velocity range of SiO emission relative to that of H$^{13}$CO$^+$, which indicates a clear association of the former with high-velocity gas. In the case of CH$_3$OH, due to blending with two other methanol lines, we employed the same velocity range as for H$^{13}$CO$^+$. This includes only the central velocity channels and excludes the emission from the line wings.%Even so, the CH$_3$OH map in Fig.\,\ref{fsptmap2} may still be slightly contaminated by emission from the blended lines.

A first glance at the SiO, H$^{13}$CO$^+$, and CH$_3$OH maps immediately reveals how differently these three molecules trace the region. While H$^{13}$CO$^+$ mostly traces the central area of NGC\,2264-C, where the gas is denser (see Peretto et al.~\citealp{peretto06}), SiO primarily emits in a more peripheral area and appears relatively depleted at the center. CH$_3$OH, on the other hand, is enhanced both at the center and at the periphery. This suggests that, in this source, CH$_3$OH traces both dense gas (as H$^{13}$CO$^+$) and shocked regions (as SiO).

\begin{deluxetable}{lcccccccc}
\tabletypesize{\footnotesize}
\tablewidth{0pt}
\tablecolumns{8}
\tablecaption{Velocity ranges and RMS noise of the integrated maps\label{tmaps}}
\tablehead{
\colhead{}  & \multicolumn{2}{c}{All channels above 5$\sigma$} & \multicolumn{2}{c}{Central velocities} & \multicolumn{4}{c}{Outflow blue- and red-shifted wings}\\
\colhead{Tracer}  & \colhead{$\Delta V_\mathrm{tot}$}  & \colhead{$\sigma_\mathrm{tot}$} & \colhead{$\Delta V_\mathrm{ctr}$\tablenotemark{a}}  & \colhead{$\sigma_\mathrm{ctr}$} & \colhead{$\Delta V_\mathrm{b}$}  & \colhead{$\sigma_\mathrm{b}$} & \colhead{$\Delta V_\mathrm{r}$} & \colhead{$\sigma_\mathrm{r}$}\\
\colhead{}  & \colhead{(km\,s$^{-1}$)}  & \colhead{(K\,km\,s$^{-1}$)} & \colhead{(km\,s$^{-1}$)}  & \colhead{(K\,km\,s$^{-1}$)} & \colhead{(km\,s$^{-1}$)}  & \colhead{(K\,km\,s$^{-1}$)} & \colhead{(km\,s$^{-1}$)} & \colhead{(K\,km\,s$^{-1}$)}}
\startdata
H$^{13}$CO$^{+}$ & [5.5,10.5] & 0.046 & [5.5,10,5] & 0.046 & --- & --- & --- & ---\\
SiO & [--1.3,20.3] & 0.12 & [5.5,10.5] & 0.087 & [--1.3,5.5] & 0.099 & [10.5,20.3] & 0.12\\
CH$_3$OH\tablenotemark{b} & --- & --- & [5.5,10.5] & 0.062 & --- & --- & --- & ---\\
CO\tablenotemark{c} & [--1.0,21.0] & 1.1 & --- & --- & [--1.0,5.5] & 0.84 & [10.5,21.0] & 1.0\\
\enddata
\tablenotetext{a}{The velocity range is determined by the channels with emission above $5\sigma$ in the H$^{13}$CO$^+$ line}
\tablenotetext{b}{The line wings are blended with other methanol lines (see Table\,\ref{tline})}
\tablenotetext{c}{The central velocities are not indicated nor used throughout the paper due to heavy self-absorption}
\end{deluxetable}

\subsection{Three SiO components} \label{3comp}

\begin{figure}
\epsscale{0.7}
%\epsscale{1}
\plotone{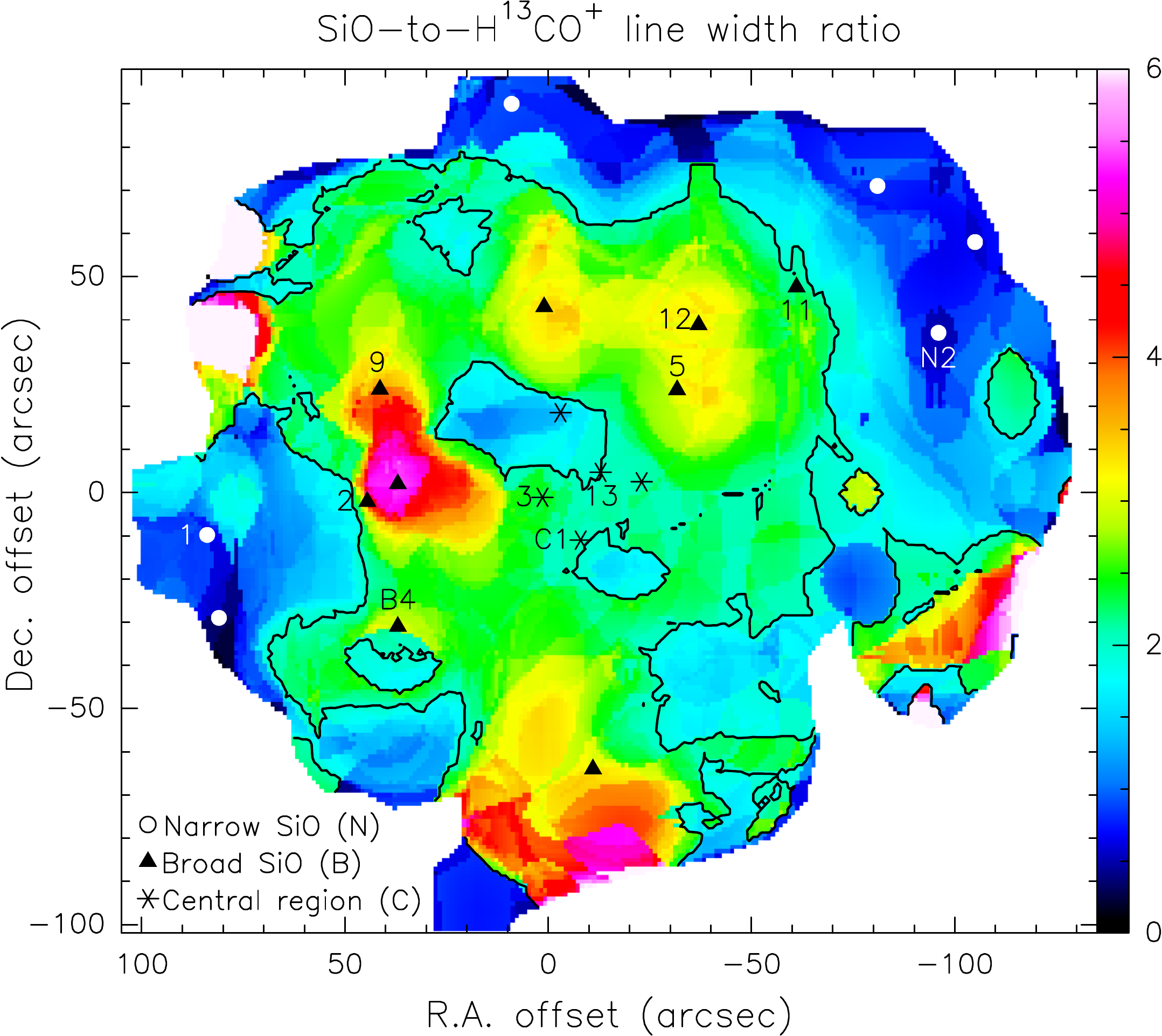}
\caption{SiO-to-H$^{13}$CO$^+$ moment 2 map within the area where the SiO emission exceeds S/N\,=\,3. The black contour marks our adopted dividing line between the narrow and broad SiO line profiles, corresponding an SiO-to-H$^{13}$CO$^+$ line width ratio of 2. The positions where we have extracted spectra for the analysis are marked with white circles (narrow component), black triangles (broad component), and asterisks (central component; see Sect.\,\ref{3comp} for details). The numbers indicate the labels of the millimeter cores as reported by Peretto et al. (2006, 2007).
\label{fratio}}
\end{figure}

In order to understand the spatial distribution and nature of the SiO emission in NGC\,2264-C, we examined both its line intensity and width, relative to those of H$^{13}$CO$^+$. To this aim, Fig.\,\ref{fratio} presents two images: the SiO-to-H$^{13}$CO$^+$ integrated map (\textit{left}), and the SiO-to-H$^{13}$CO$^+$ moment\,2 --or line width-- map (\textit{right}). The former was obtained using the same velocity interval for both SiO and H$^{13}$CO$^+$ ($\Delta V_\mathrm{ctr}$ in Table\,\ref{tmaps}), while for the latter we used the entire velocity ranges with emission above 5$\sigma$ ($\Delta V_\mathrm{tot}$ in Table\,\ref{tmaps}). The plotted area comprises all the positions where the emission of SiO exceeds a S/N of 3. Within this area, the S/N of the H$^{13}$CO$^+$ emission always exceeds 15.

Figure\,\ref{fsptmap1} and \ref{fratio} (\textit{left}) clearly show how the strongest SiO emission is offset with respect to the massive molecular core CMM3 at the center of NGC\,2264-C, thus tracing a relatively peripheral area of the molecular clump. From Fig.\,\ref{fratio} (\textit{right}), one notices that the width of the SiO line also tends to be broad in this area, more than twice that of H$^{13}$CO$^+$. This region is thus characterized by high-velocity shocks that are likely the result of a complex network of protostellar outflows within NGC\,2264-C. Close to CMM3, the SiO line emission is weaker but nevertheless broad, also suggesting the presence of high-velocity shocks. Finally, further out with respect to the enhanced SiO emission region, there is an area where SiO lines are both relatively weak and narrow, especially towards the north and north-west. Indeed, the line width of SiO in this external area is comparable to that of H$^{13}$CO$^+$. Thus, overall, we can distinguish three different SiO components:

\begin{itemize}
\item Broad-line SiO component (hereafter \textit{broad} or B): it comprises the area where the emission of SiO is most prominent. The SiO lines have full-widths at zero power (FWZP) between 20 and 40\,km\,s$^{-1}$. This component is spatially offset with respect to the center of the molecular clump, which, together with the high-velocity SiO line wings associated with it, indicates the presence of a circumcluster shocked region due to protostellar activity within the molecular clump.
\item Central SiO component (hereafter \textit{central} or C): it lies at the center of the clump, around the molecular cores CMM3 and CMM13, where the gas is densest according to Peretto et al. \cite{peretto06}. It is characterized by weaker SiO emission than the broad component, but it also displays high-velocity line wings suggestive of protostellar outflows, with FWZP typically between 20 and 25\,km\,s$^{-1}$.
\item Narrow-line SiO component (hereafter \textit{narrow} or N): it is located at the very periphery of NGC\,2264-C, and it exhibits weak and narrow ($\sim\,2\,$km\,s$^{-1}$) SiO lines relative to H$^{13}$CO$^+$.
\end{itemize}

\begin{figure*}
\epsscale{0.342}
%\epsscale{0.684}
\plotone{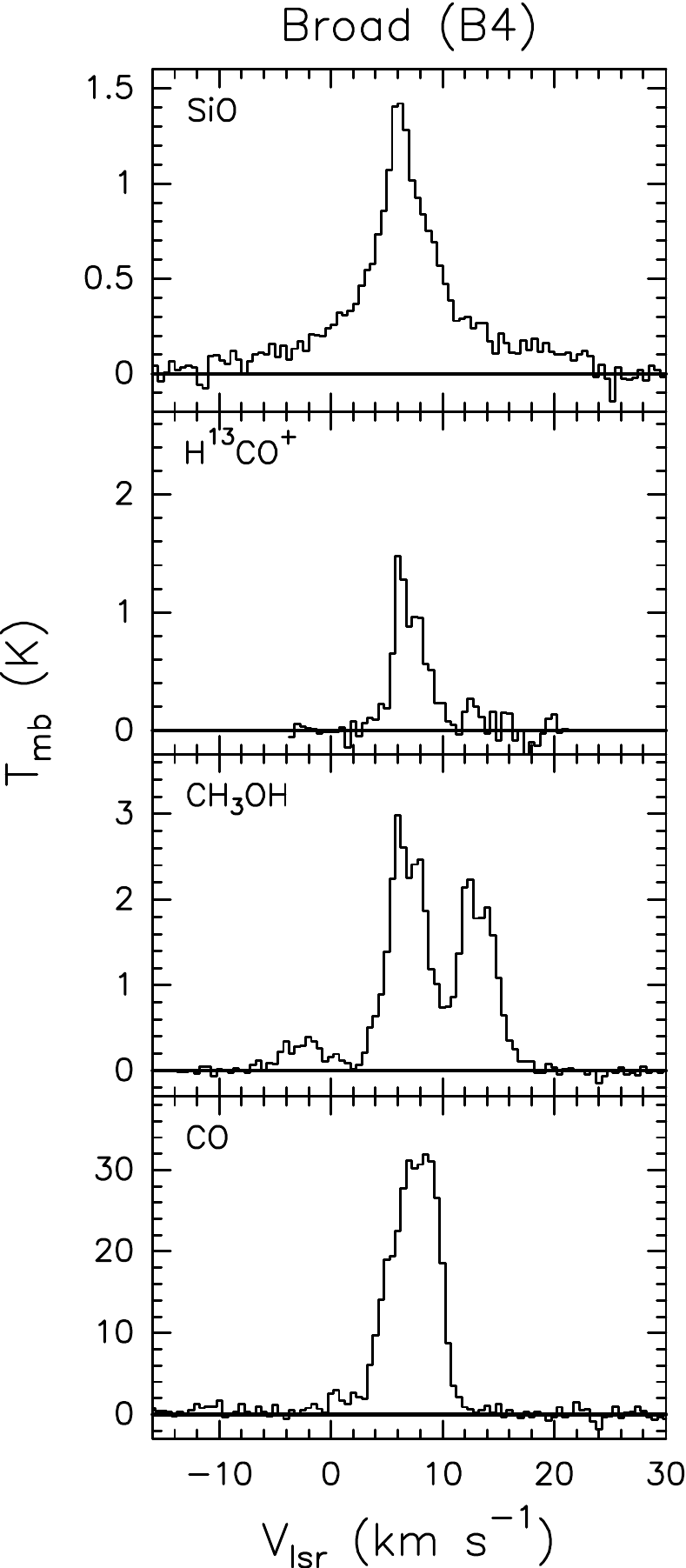}
\epsscale{0.3}
%\epsscale{0.6}
\plotone{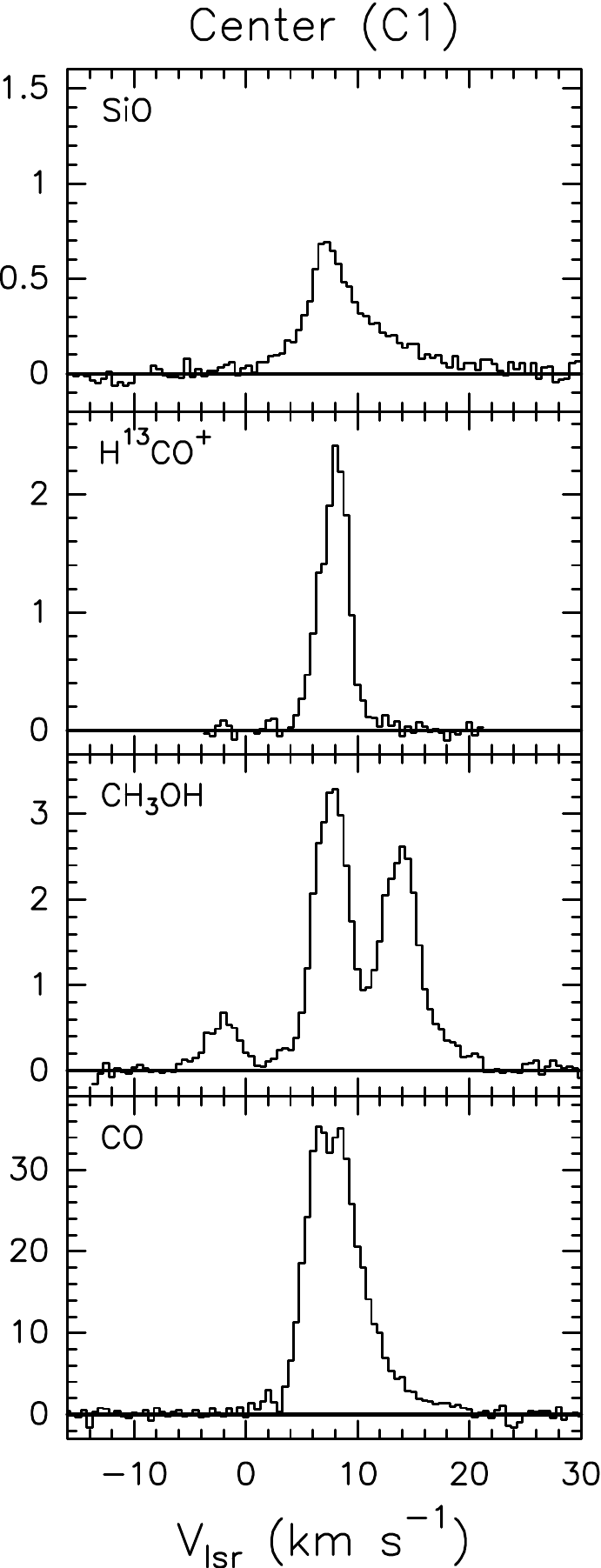}
\plotone{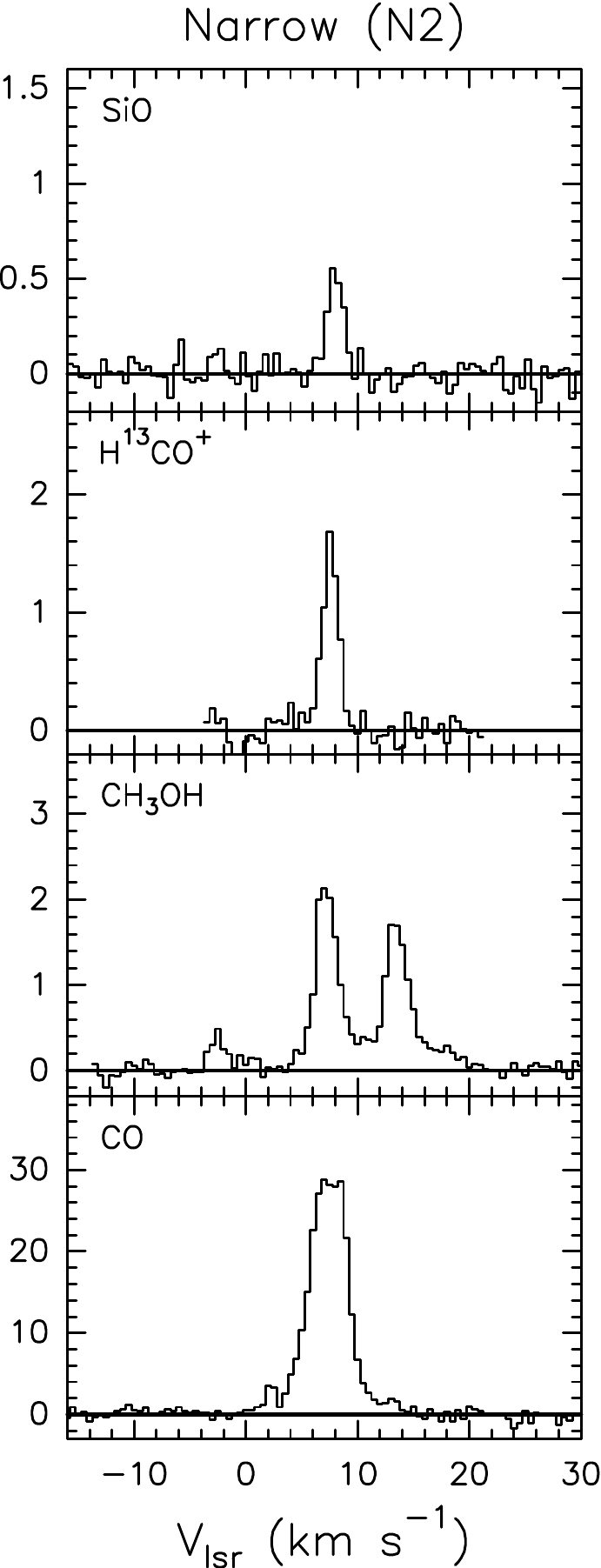}
\caption{Sample spectra of CO, CH$_3$OH, H$^{13}$CO$^+$, and SiO, extracted from the offset positions of B4, C1, and N2 (see Table\,\ref{tflux} and labels in Fig.\,\ref{fratio}), illustrating, respectively, the three components that we distinguish: broad (\textit{left}), central (\textit{middle}), and narrow (\textit{right}), as described in the main text. The average systemic velocity across NGC\,2264-C is 7.5\,km\,s$^{-1}$.
\label{fsptsample}}
\end{figure*}

For reference, Fig.\,\ref{fsptsample} shows sample spectra for each of the three different components, extracted at the positions indicated in the figure caption and labelled in Fig.\,\ref{fratio} (B4, C1, and N2). Following the above classification, we have selected and marked, in Fig.\,\ref{fratio}, several positions representative of each of the three SiO components, whose molecular spectra are used in Sect.\,\ref{derivation} to derive their respective molecular column densities and abundances. Some of these coincide with molecular cores identified by Peretto et al.~(\citealp{peretto06}, \citealp{peretto07}). For example, CMM3 and CMM13 are two of the selected positions representing the central component. Table\,\ref{tflux} lists the offset coordinates, velocity ranges with emission above S/N\,=\,3, and line fluxes measured at each of these positions. The numbers in parentheses are the measured flux errors, which account for the RMS noise of the observations and the calibration uncertainty.

\begin{deluxetable}{lcccccc}
\tabletypesize{\footnotesize}
\tablewidth{0pt}
\tablecolumns{7}
\tablecaption{Velocity ranges and line fluxes measured in each selected position\label{tflux}}
\tablehead{
\colhead{Label$\tablenotemark{a}$}  & \colhead{$(\Delta x,\Delta y)$\tablenotemark{b}}  & \colhead{$\Delta V$} & \colhead{$\int T_\mathrm{mb} dV$\,[H$^{13}$CO$^+$]\tablenotemark{c}}  & \colhead{$\int T_\mathrm{mb} dV$\,[SiO]} & \colhead{$\int T_\mathrm{mb} dV$\,[CH$_3$OH]\tablenotemark{d}}  & \colhead{$\int T_\mathrm{mb} dV$\,[CO]}\\
\colhead{}  & \colhead{($''$,$''$)} & \colhead{(km\,s$^{-1}$)} & \colhead{(K\,km\,s$^{-1}$)} &  \colhead{(K\,km\,s$^{-1}$)} & \colhead{(K\,km\,s$^{-1}$)}  & \colhead{(K\,km\,s$^{-1}$)}}
\startdata
\cutinhead{Narrow component (N)}
 CMM1 (ctr) & (83.9,--9.8) & [5.2,11.2] &  2.7 (0.8) &  0.93 (0.29) &  4.6 (1.4) & --- \\
   N1 (ctr) & (9.0,90.0) & [5.2,9.2] &  2.2 (0.7) &  0.58 (0.19) &  3.2 (0.9) & --- \\
   N2 (ctr) & (--96.0,37.0) & [6.2,8.8] &  2.7 (0.8) &  0.72 (0.22) &  4.4 (1.3) & --- \\
   N3 (ctr) & (--105.0,58.0) & [6.8,9.8] &  2.5 (0.7) &  0.70 (0.22) &  3.4 (1.0) & ---\\
   N4 (ctr) & (81.0,--29.0) & [4.8,11.8] &  3.4 (1.0) & 1.1 (0.3) &  3.8 (1.1) & ---\\
   N5 (ctr) & (--81.0,71.0) & [6.5,10.0] &  3.0 (0.9) &  0.97 (0.30) &  4.6 (1.4) & ---\\
\cutinhead{Broad component (B)}
 CMM2 (ctr) & (44.5,--2.1) & [4.8,8.2] &  4.0 (1.2) & 1.9 (0.6) &  7.0 (2.1) & ---\\
 CMM2 (blue) &  & [--1.2,4.8] & --- &  1.1 (0.3) & --- & 34 (10)\\
 CMM2 (red) &  & [8.2,22.8] & --- &  2.6 (0.8) & --- & 60 (18)\\[2mm]
 CMM5 (ctr) & (--31.7,23.8) & [5.2,10.8] &  7.8 (2.3) &  4.6 (1.4) & 15 (4) & --- \\
 CMM5 (blue) &  & [0.8,5.2] & --- &  1.1 (0.3) & --- & 40 (12)\\
 CMM5 (red) &  & [10.8,20.8] & --- &  3.1 (0.9) & --- & 61 (18)\\[2mm]
 CMM9 (ctr) & (41.4,23.9) & [4.8,8.8] &  4.3 (1.3) & 3.1 (0.9) &  8.6 (2.6) & ---\\
 CMM9 (blue) &  & [--9.2,4.8] & --- &  3.2 (1.0) & --- & 22 (7)\\
 CMM9 (red) &  & [8.8,28.8] & --- &  3.2 (1.0) & --- & 34 (10)\\[2mm]
CMM11 (ctr) & (--61.1,47.6) & [5.8,9.8] &  6.0 (1.8) & 2.8 (0.8) & 8.5 (2.6) & ---\\
CMM11 (blue) &  & [1.8,5.8] & --- &  0.76 (0.23) & --- & 40 (12)\\
CMM11 (red) &  & [9.8,25.8] & --- &  2.3 (0.7) & --- & 33 (10)\\[2mm]
CMM12 (ctr) & (--37.0,38.8) & [5.2,11.2] &  7.6 (2.3) &  7.1 (2.1) & 16 (5) & ---\\
CMM12 (blue) &  & [--5.2,5.2] & --- &  2.1 (0.6) & --- & 43 (13)\\
CMM12 (red) &  & [11.2,25.8] & --- &  3.9 (1.2) & --- & 43 (13)\\[2mm]
   B1 (ctr) & (37.0,2.0) & [4.8,7.8] &  4.2 (1.3) &  1.8 (0.5) &  7.1 (2.1) & ---\\
   B1 (blue) &  & [--1.2,4.8] & --- &  1.2 (0.4) & --- & 23 (7)\\
   B1 (red) &  & [7.8,22.8] & --- &  3.6 (1.1) & --- & 73 (22)\\[2mm]
   B2 (ctr) & (--11.0,--64.0) & [5.8,10.2] &  2.8 (0.8) &  2.5 (0.7) & 11 (3) & ---\\
   B2 (blue) &  & [--1.2,5.8] & --- &  0.63 (0.20) & --- & 18 (5)\\
   B2 (red) &  & [10.2,24.8] & --- &  5.8 (1.7) & --- & 47 (14)\\[2mm]
   B3 (ctr) & (1.0,43.0) & [5.2,9.8] &  4.5 (1.4) &  3.9 (1.2) &  9.4 (2.8) & ---\\
   B3 (blue) &  & [0.8,5.2] & --- &  1.0 (0.3) & --- & 17 (5)\\
   B3 (red) &  & [9.8,20.8] & --- &  3.6 (1.1) & --- & 43 (13)\\[2mm]
   B4 (ctr) & (37.0,--31.0) & [5.2,9.2] &  3.6 (1.1) &  3.8 (1.1) &  9.0 (2.7) & --- \\
   B4 (blue) &  & [--6.5,5.2] & --- &  3.9 (1.2) & --- & 35 (11)\\
   B4 (red) &  & [9.2,23.5] & --- &  3.1 (0.9) & --- & 33 (10)\\
\cutinhead{central component (C)}
 CMM3 (ctr) & (1.4,--1.1) & [4.8,10.2] &  7.9 (2.4) &  3.2 (1.0) & 13 (4) & ---\\
 CMM3 (blue) &  & [--2.2,4.8] & --- &  0.95 (0.29) & --- & 16 (5)\\
 CMM3 (red) &  & [10.2,22.8] & --- &  2.0 (0.6) & --- & 31 (9)\\[2mm]
CMM13 (ctr) & (--12.9,4.7) & [4.8,10.8] &  7.8 (2.3) &  3.0 (0.9) & 14 (4) & ---\\
CMM13 (blue) &  & [--1.2,4.8] & --- &  0.50 (0.16) & --- & 16 (5)\\
CMM13 (red) &  & [10.8,17.8] & --- &  1.1 (0.3) & --- & 32 (10)\\[2mm]
  C1 (ctr) & (--8.0,--11.0) & [4.8,10.8] &  7.1 (2.1) &  2.7 (0.8) & 13 (4) & ---\\
   C1 (blue) &  & [1.8,4.8] & --- &  0.43 (0.14) & --- & 13 (4)\\
   C1 (red) &  & [10.8,18.8] & --- &  1.2 (0.4) & --- & 36 (11)\\[2mm]
  C2 (ctr) & (--3.0,18.5) & [5.2,9.8] &  6.3 (1.9) &  2.9 (0.9) & 12 (4) & ---\\
   C2 (blue) &  & [--1.2,5.2] & --- &  0.84 (0.26) & --- & 27 (8)\\
   C2 (red) &  & [9.8,17.8] & --- &  1.4 (0.4) & --- & 43 (13)\\[2mm]
   C3 (ctr) & (--23.0,2.5) & [5.2,10.8] &  7.4 (2.2) &  2.7 (0.8) & 13 (4) & --- \\
   C3 (blue) &  & [--0.2,5.2] & --- &  0.52 (0.17) & --- & 23 (7)\\
   C3 (red) &  & [10.8,18.8] & --- &  1.4 (0.4) & --- & 52 (16)\\[2mm]
\enddata
\tablenotetext{a}{The label in brackets indicates whether the velocity range used corresponds to the central velocities (ctr), the blue-shifted wing (blue) or the red-shifted wing (red)}
\tablenotetext{b}{Offset coordinates with respect to the reference position: R.A.(J2000)\,=\,06$^\mathrm{h}$41$^\mathrm{m}$12.3$^\mathrm{s}$, Dec.(J2000)\,=\,09$^\circ$29$'$11.90$''$}
\tablenotetext{c}{High-velocity emission is not detected for H$^{13}$CO$^+ (J = 1-0)$}
\tablenotetext{d}{No line flux measured at the line wings due to blending with two other CH$_3$OH lines}
\end{deluxetable}

\subsection{Outflow maps}\label{out}

\begin{figure*}
\epsscale{0.51}
%\epsscale{1.02}
\plotone{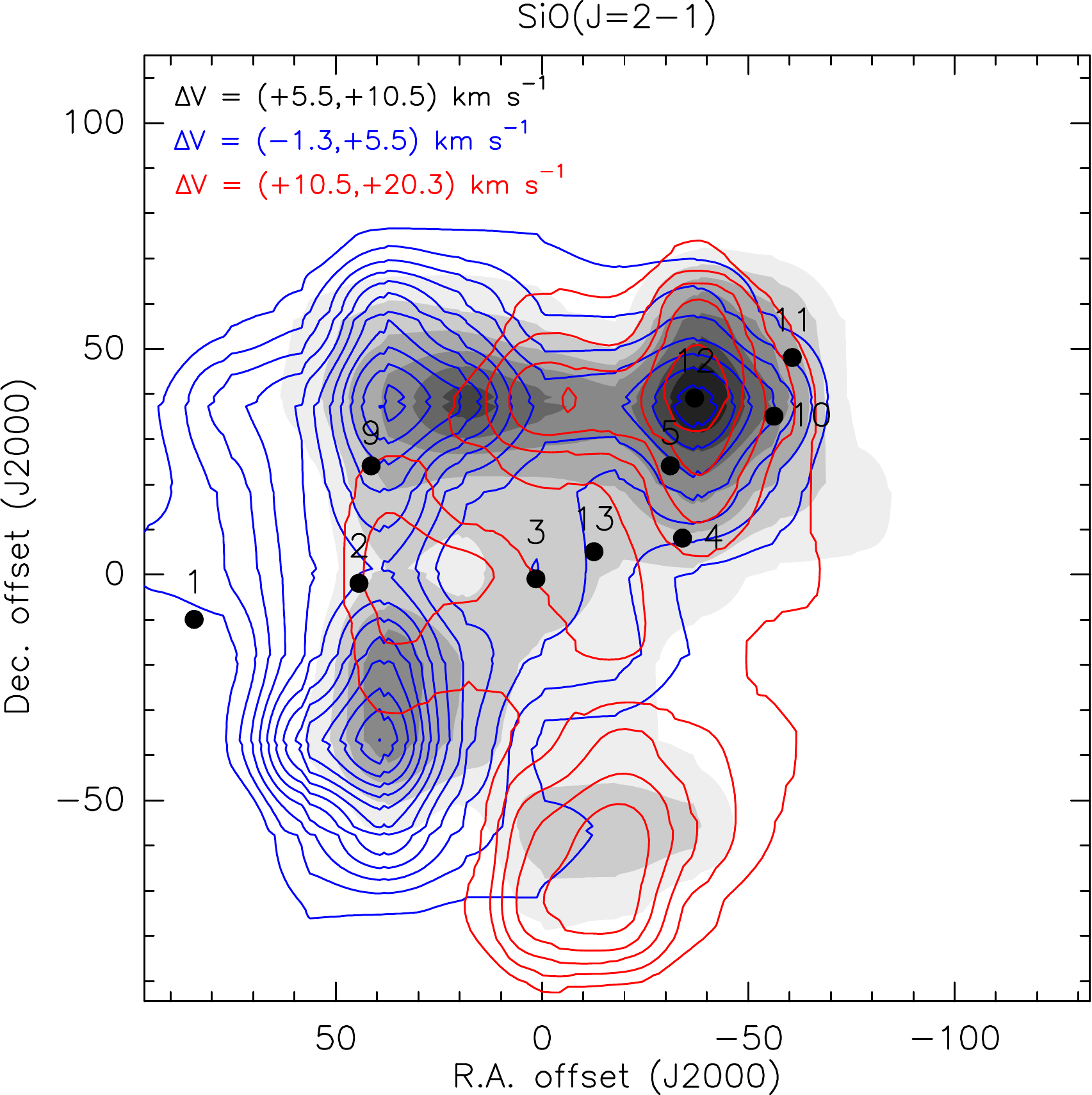}
\epsscale{0.415}
%\epsscale{0.83}
\plotone{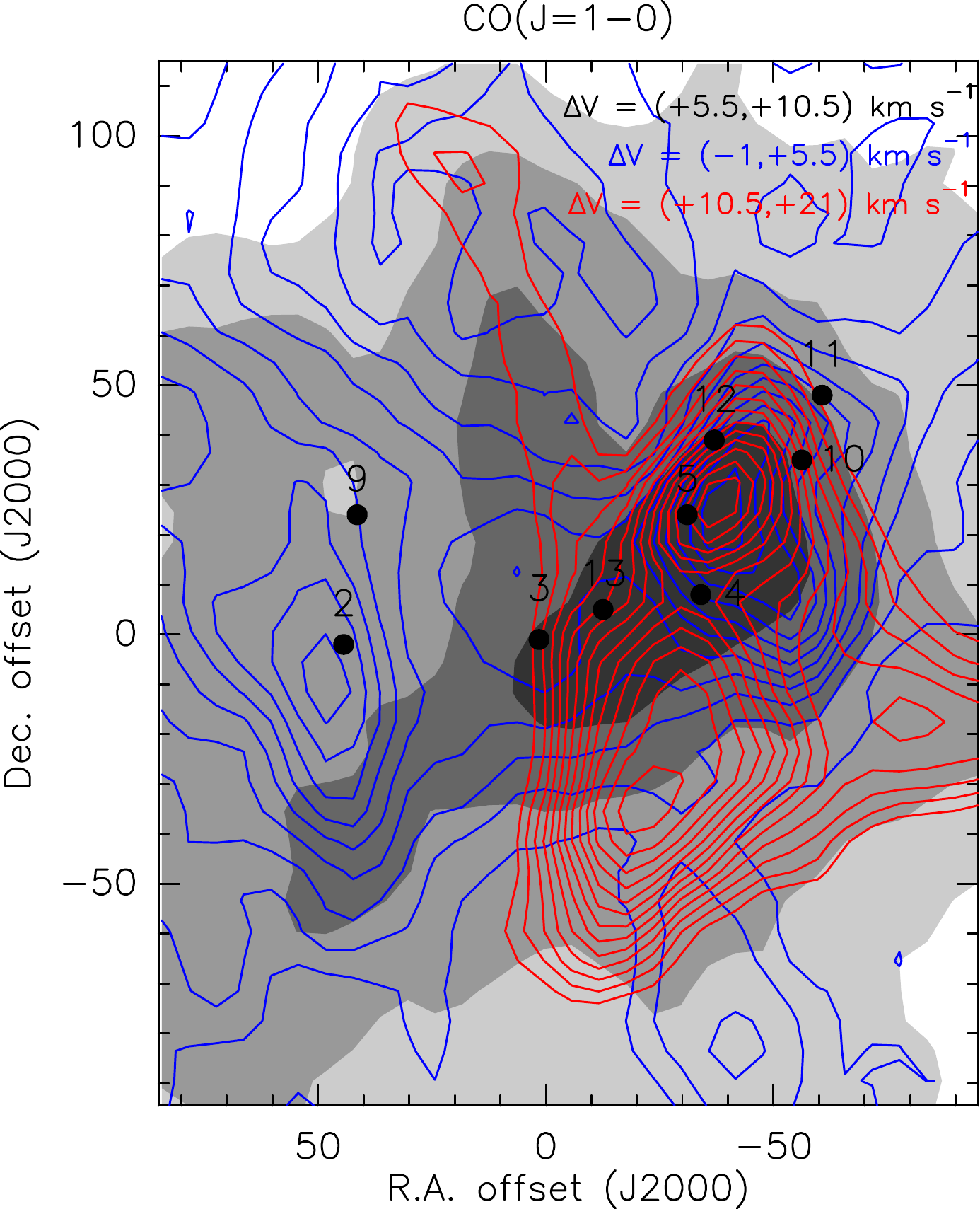}
\caption{SiO and CO outflow maps. Blue and red contours correspond to the blue- and red-shifted high velocity wings, as indicated in the key and in Table\,\ref{tmaps}. The gray scale corresponds to the central velocities, as indicated in the key, with contours starting at 30\% (SiO) and 50\% (CO) of the peak intensity and increasing by steps of 10\%. The peak intensity is 6.7\,K\,km\,s$^{-1}$ and 172\,K\,km\,s$^{-1}$ for the SiO and CO gray scales, respectively. Contours for the SiO outflow maps start at 5$\sigma$ and increase by steps of 3$\sigma$, while for the CO maps they start at 25$\sigma$ and increase by steps of 5$\sigma$ (see Table\,\ref{tmaps}). Black circles mark the positions of the cores identified by Peretto et al. (\citealp{peretto06}, \citealp{peretto07}).
\label{fout}}
\end{figure*}

High-velocity SiO emission is widely present across $\sim 0.5$\,pc in NGC\,2264-C. This is strongly supportive of a large amount of protostellar outflow activity, as reported by Maury et al. \cite{maury09} from CO($J = 2-1$) observations. We have obtained SiO($J = 2-1$) and CO($J = 1-0$) outflow maps by integrating the blue- and red-shifted high-velocity emission of the lines. These are presented in Fig.\,\ref{fout}.

Since CH$_3$OH is blended with other methanol lines, an outflow assessment from this molecular tracer is unreliable. Nevertheless, a qualitative analysis of the unblended blue- and red-shifted line wings from the two E-state transitions of CH$_3$OH (see the spectra in Fig.\,\ref{fsptsample}) shows that they are not as prominent as those of SiO, which span more than 10\,km\,s$^{-1}$ from the ambient velocities. The outflow maps using the unblended line wings of methanol show a similar distribution to that of SiO (Fig.\,\ref{fout}). However, since this emission arises from two lines that are not considered in the analysis below, we do not present the maps here. For CO($J = 1-0$), the line wings are not so extended either, but this is most likely due to the much lower sensitivity --by a factor 10-- of the observations for this molecule (see Table\,\ref{tline}).

The outflow maps in Fig.\,\ref{fout} show how the spatial distribution of the blue- and red-shifted high-velocity emission is rather complex. The blue-shifted emission is dominant towards the east of the clump, while the red-shifted emission concentrates mostly towards the west. Unfortunately, the angular resolution of our observations does not allow us to clearly separate the individual molecular outflows and associate them to particular molecular cores. The maps rather show a tangled circumcluster shock network. Despite this, hints of the very collimated molecular outflow defined by Maury et al. \cite{maury09} as flows F1 (red-shifted; north of CMM13) and F2 (blue-shifted; south of CMM13), are evident in our CO outflow map.  It should be noticed that the CO emission may be moderately optically thick even at the high-velocity wings (see Sect.\,\ref{high}). 

\section{Derivation of column densities and abundances}\label{derivation}

In order to evaluate the nature of the different SiO components identified in Sect.\,\ref{3comp}, we provide here an estimate of molecular column densities and abundances at each selected position (Table\,\ref{tflux}).

\subsection{Column density of the narrow component}\label{low}

Under the assumption of Local Thermodynamic Equilibrium (LTE), the total column density of a given molecule, $N$, is described by Boltzmann's equation as:

\begin{equation}
N = \frac{N_u}{g_u} Q(T_\mathrm{ex}) e^{-\frac{E_u}{k T_\mathrm{ex}}}
\label{entot}
\end{equation}
where $N_u$, $g_u$, $T_\mathrm{ex}$, $Q(T_\mathrm{ex})$, $E_u$, and $k$ are, respectively, the column density of the upper energy-state, its degeneracy, the excitation temperature describing the level populations, the partition function at a given $T_\mathrm{ex}$, the energy of the upper state, and Boltzmann's constant.

The column density of the upper-energy state is given by:

\begin{equation}
N_u = \frac{8 \pi \nu^3}{c^3} \frac{1}{A_{ul}} \frac{1}{e^\frac{h \nu}{k T_\mathrm{ex}} - 1} \int \tau dv
\label{enu}
\end{equation}
where $\nu$, $c$, $A_{ul}$, $h$ and $\int \tau dv$ are the frequency of the transition, speed of light, Einstein coefficient for spontaneous emission, Planck's constant, and the velocity-integrated optical depth of the molecular line, respectively. With only one observed transition per molecule, we have no means to constrain $T_\mathrm{ex}$. We therefore adopt three different values: $T_\mathrm{ex} = 10$, 20, and 40\,K. For each of these, and for each molecular tracer, we have estimated the integrated optical depth as the discrete summation of the optical depth at each velocity channel, $n$, of constant width $\delta v$:

\begin{equation}
\sum_{n} \tau_n \delta v = -\delta v \sum_n \ln(1 - \frac{T_\mathrm{mb}}{J_\nu(T_\mathrm{ex}) - J_\nu(T_\mathrm{bg})})
\label{etau}
\end{equation}
Here $T_\mathrm{mb}$ is the main beam temperature of the line for a given channel $n$, and $J_\nu(T)$ is defined as:

\begin{equation}
J_\nu(T) = \frac{h \nu}{k} \frac{1}{e^\frac{h \nu}{k T} - 1}
\label{ej}
\end{equation}

Table\,\ref{tflux} lists the velocity ranges used to integrate the emission and obtain the column density of SiO, CH$_3$OH, and H$^{13}$CO$^+$ in the six positions selected to represent the narrow component. These ranges comprise the velocity channels where the emission of H$^{13}$CO$^+$, which displays no high-velocity wings, exceeds 3$\sigma$. In the case of CMM1, the emission is double peaked, and hence a wide velocity range is measured.

The resulting column densities for each of the three adopted values of $T_\mathrm{ex}$ are presented in Table\,\ref{tcoln}. Since the central velocities of CO($J = 1-0$) correspond to heavily self-absorbed emission in most of the mapped area, we do not consider them here. Instead, we will analyse the emission of the CO high-velocity line wings in Sect.\,\ref{high}.

\begin{deluxetable}{lcccccc}
\tabletypesize{\footnotesize}
\tablewidth{0pt}
\tablecolumns{7}
\tablecaption{Narrow component: column densities and abundances in each selected position\label{tcoln}}
\tablehead{
\colhead{Label}  & \colhead{$T_\mathrm{ex}$}  & \colhead{$N_\mathrm{SiO}$}  & \colhead{$N_\mathrm{CH3OH}$} & \colhead{$N_\mathrm{H13CO+}$} & \colhead{$\chi_\mathrm{SiO}$} & \colhead{$\chi_\mathrm{CH3OH}$}\\
\colhead{}  & (K) & \colhead{($10^{12}$\,cm$^{-2}$)} & \colhead{($10^{14}$\,cm$^{-2}$)} & \colhead{($10^{12}$\,cm$^{-2}$)} & \colhead{($10^{-11}$)} & \colhead{($10^{-9}$)}}
\startdata
 CMM1 & 10 & $2.1 \pm 0.8$ & $4.0 \pm 1.2$ & $3.2 \pm 1.2$ & $5.3 \pm2.8$ & $10 \pm 5$\\
  & 20 & $2.7 \pm 1.0$ & $7.2 \pm 2.2$ & $4.5 \pm 1.6$ & $5.0 \pm 2.6$ & $13 \pm 6$\\
  & 40 & $4.4 \pm 1.7$ & $17 \pm 5$ & $7.6 \pm 2.8$ & $4.8 \pm 2.5$ & $19 \pm 9$\\
   N1 & 10 & $1.3 \pm 0.6$ & $2.8 \pm 0.9$ & $2.7 \pm 0.9$ & $4.0 \pm 2.4$ & $8.8 \pm 4.1$\\
  & 20 & $1.7 \pm 0.8$ & $5.0 \pm 1.6$ & $3.7 \pm 1.3$ & $3.8 \pm 2.2$ & $11 \pm 5$\\
  & 40 & $2.7 \pm 1.3$ & $12 \pm 4$ & $6.3 \pm 2.3$ & $3.6 \pm 2.2$ & $16 \pm 8$\\
   N2 & 10 & $1.6 \pm 0.6$ & $4.1 \pm 1.2$ & $3.4 \pm 1.1$ & $3.9 \pm 1.8$ & $10 \pm 4$\\
  & 20 & $2.1 \pm 0.7$ & $7.1 \pm 2.1$ & $4.7 \pm 1.5$ & $3.7 \pm 1.8$ & $13 \pm 5$\\
  & 40 & $3.4 \pm 1.2$ & $17 \pm 5$ & $7.8 \pm 2.5$ & $3.6 \pm 1.7$ & $18 \pm 8$\\
   N3 & 10 & $1.6 \pm 0.6$ & $3.1 \pm 0.9$ & $3.1 \pm 0.9$ & $4.3 \pm 2.1$ & $8.4 \pm 3.6$\\
  & 20 & $2.0 \pm 0.8$ & $5.5 \pm 1.6$ & $4.2 \pm 1.3$ & $4.1 \pm 2.0$ & $11 \pm 5$\\
  & 40 & $3.3 \pm 1.3$ & $13 \pm 4$ & $7.0 \pm 2.2$ & $3.9 \pm 2.0$ & $15 \pm 7$\\   
   N4 & 10 & $2.3 \pm 1.1$ & $3.4 \pm 1.0$ & $4.1 \pm 1.3$ & $4.8 \pm 2.7$ & $7.0 \pm 3.1$\\
  & 20 & $3.0 \pm 1.4$ & $6.1 \pm 1.8$ & $5.7 \pm 1.9$ & $4.5 \pm 2.6$ & $9.0 \pm 4.0$\\
  & 40 & $5.0 \pm 2.3$ & $15 \pm 4$ & $9.6 \pm 3.2$ & $4.3 \pm 2.5$ & $13 \pm 6$\\
   N5 & 10 & $2.2 \pm 0.8$ & $4.4 \pm 1.2$ & $3.9 \pm 1.2$ & $4.6 \pm 2.2$ & $9.3 \pm 3.8$\\
  & 20 & $2.8 \pm 1.0$ & $7.5 \pm 2.2$ & $5.3 \pm 1.7$ & $4.4 \pm 2.1$ & $12 \pm 5$\\
  & 40 & $4.5 \pm 1.6$ & $18 \pm 5$ & $8.8 \pm 2.8$ & $4.3 \pm 2.1$ & $17 \pm 7$\\
\enddata
\end{deluxetable}

\subsection{Column density of the broad and central components}\label{high}

Following an analogous procedure to that described in Sect.\,\ref{low}, we derived the SiO and CO column densities at each selected position of the broad and central components, which trace high-velocity molecular gas from the protocluster activity. For CO, we integrated the line emission only at the blue- and red-shifted line wings, as indicated in column\,3 of Table\,\ref{tflux}. The velocity ranges used in the integration were determined by comparing the CO line with the H$^{13}$CO$^+$ line and following the same approach described in L\'opez-Sepulcre et al. \cite{ls09}: we defined the low-velocity limits where the line intensity of H$^{13}$CO$^+$ falls below 2$\sigma$, and the high-velocity limits where the line intensity of the two outflow tracers falls below 2$\sigma$. For SiO, we provide two estimates of the column density: one that includes only the blue- and red-shifted line wings ($N_\mathrm{SiO}$ (blue) and $N_\mathrm{SiO}$ (red)), as in the case of CO, and one that includes the total column density of SiO integrated across the entire velocity range where the emission exceeds $2\sigma$ ($N_\mathrm{SiO}$ (tot)). 

The resulting column densities, again adopting $T_\mathrm{ex} = 10$, 20, and 40\,K, are listed in Table\,\ref{tcolb}. We note that, while the SiO emission is mostly optically thin across the whole velocity range, that of CO is moderately optically thick even in the high-velocity wings. The velocity-averaged optical depth in the CO wings ranges between 0.04 and 3.9 depending on the offset position and the excitation temperature adopted, with typical values of 0.12 ($T_\mathrm{ex} = 40$\,K) and 0.9 ($T_\mathrm{ex} = 10$\,K). These optical depths are taken into account in the derivation of column densities, as detailed in Sect.\,\ref{low}.

\begin{deluxetable}{lcccccccc}
\tabletypesize{\footnotesize}
\tablewidth{0pt}
\tablecolumns{9}
\tablecaption{Broad and central components: column densities and abundances in each selected position\label{tcolb}}
\tablehead{
\colhead{Label}  & \colhead{$T_\mathrm{ex}$} & \colhead{$N_\mathrm{SiO}$ (tot)\tablenotemark{a}}  & \colhead{$N_\mathrm{SiO}$ (blue)} & \colhead{$N_\mathrm{SiO}$ (red)} & \colhead{$N_\mathrm{CO}$ (blue)\tablenotemark{b}} & \colhead{$N_\mathrm{CO}$ (red)} & $\chi_\mathrm{SiO}$ (blue) & $\chi_\mathrm{SiO}$ (red)\\
\colhead{}  & (K) & \colhead{($10^{12}$\,cm$^{-2}$)} & \colhead{($10^{12}$\,cm$^{-2}$)} & \colhead{($10^{12}$\,cm$^{-2}$)} & \colhead{($10^{16}$\,cm$^{-2}$)} & \colhead{($10^{16}$\,cm$^{-2}$)} & \colhead{($10^{-9}$)} & \colhead{($10^{-9}$)}}
\startdata
\cutinhead{Broad component (B)}
 CMM2 & 10 & $13 \pm 3$ & $2.5 \pm 0.9$ & $5.8 \pm 2.1$ & $6.8 \pm 1.3$ & $8.7 \pm 2.1$ & $3.8 \pm 1.5$ & $6.7 \pm 2.8$\\
  & 20 & $16 \pm 3$ & $3.3 \pm 1.2$ & $7.6 \pm 2.7$ & $5.1 \pm 1.4$ & $8.7 \pm 2.5$ & $6.4 \pm 2.9$ & $8.7 \pm 4.0$\\
  & 40 & $27 \pm 6$ & $5.4 \pm 1.9$ & $12 \pm 4$ & $7.6 \pm 2.2$ & $13 \pm 4$ & $7.1 \pm 3.3$ & $9.3 \pm 4.3$\\
 CMM5 & 10 & $20 \pm 4$ & $2.4 \pm 0.8$ & $7.0 \pm 2.2$ & --- & $16 \pm 2$ & --- & $4.4 \pm 1.6$\\
  & 20 & $26 \pm 5$ & $3.1 \pm 1.1$ & $9.0 \pm 2.9$ & $7.1 \pm 1.8$ & $9.6 \pm 2.6$ & $4.4 \pm 1.9$ & $9.4 \pm 4.0$\\
  & 40 & $42 \pm 8$ & $5.1 \pm 1.7$ & $15 \pm 5$ & $9.4 \pm 2.7$ & $14 \pm 4$ & $5.4 \pm 2.4$ & $10 \pm 5$\\
 CMM9 & 10 & $21 \pm 4$ & $7.1 \pm 2.3$ & $7.0 \pm 2.5$ & $2.3 \pm 0.8$ & $3.7 \pm 1.0$ & $31 \pm 13$ & $19 \pm 9$\\
  & 20 & $28 \pm 5$ & $9.3 \pm 3.1$ & $9.2 \pm 3.3$ & $2.9 \pm 0.9$ & $4.6 \pm 1.4$ & $32 \pm 14$ & $20 \pm 9$\\
  & 40 & $45 \pm 9$ & $15 \pm 5$ & $15 \pm 5$ & $4.7 \pm 1.4$ & $7.3 \pm 2.2$ & $32 \pm 14$ & $21 \pm 10$\\
 CMM11 & 10 & $13 \pm 3$ & $1.7 \pm 0.5$ & $5.2 \pm 1.8$ & --- & $3.6 \pm 1.0$ & --- & $14 \pm 6$\\
  & 20 & $17 \pm 3$ & $2.2 \pm 0.7$ & $6.7 \pm 2.3$ & $7.7 \pm 1.8$ & $4.4 \pm 1.3$ & $2.8 \pm 1.1$ & $15 \pm 7$\\
  & 40 & $28 \pm 6$ & $3.6 \pm 1.2$ & $11 \pm 4$ & $9.7 \pm 2.7$ & $7.0 \pm 2.1$ & $3.7 \pm 1.6$ & $16 \pm 7$\\   
 CMM12 & 10 & $31 \pm 6$ & $4.8 \pm 1.7$ & $8.7 \pm 2.9$ & $6.2 \pm 1.5$ & $5.3 \pm 1.4$ & $7.7 \pm 3.3$ & $17 \pm 7$\\
  & 20 & $39 \pm 8$ & $6.2 \pm 2.2$ & $11 \pm 4$ & $6.2 \pm 1.8$ & $6.0 \pm 1.7$ & $10 \pm 5$ & $19 \pm 8$\\
  & 40 & $62 \pm 12$ & $10 \pm 4$ & $18 \pm 6$ & $9.5 \pm 2.8$ & $9.4 \pm 2.8$ & $11 \pm 5$ & $20 \pm 9$\\
 B1 & 10 & $15 \pm 3$ & $2.6 \pm 1.1$ & $8.1 \pm 3.0$ & $3.2 \pm 0.8$ & $12 \pm 3$ & $8.2 \pm 3.9$ & $6.5 \pm 2.8$\\
  & 20 & $19 \pm 4$ & $3.4 \pm 1.4$ & $11 \pm 4$ & $3.3 \pm 0.9$ & $11 \pm 3$ & $10 \pm 5$ & $9.6 \pm 4.5$\\
  & 40 & $31 \pm 7$ & $5.6 \pm 2.3$ & $17 \pm 6$ & $5.1 \pm 1.5$ & $16 \pm 5$ & $11 \pm 6$ & $10 \pm 5$\\
 B2 & 10 & $20 \pm 5$ & $1.4 \pm 0.7$ & $13 \pm 4$ & $2.1 \pm 0.6$ & $5.9 \pm 1.5$ & $6.7 \pm 4.0$ & $22 \pm 9$\\
  & 20 & $26 \pm 6$ & $1.8 \pm 1.0$ & $17 \pm 5$ & $2.4 \pm 0.7$ & $6.5 \pm 1.9$ & $7.5 \pm 4.6$ & $26 \pm 11$\\
  & 40 & $42 \pm 10$ & $2.9 \pm 1.6$ & $27 \pm 9$ & $3.8 \pm 1.1$ & $10 \pm 3$ & $7.7 \pm 4.7$ & $27 \pm 12$\\
 B3 & 10 & $20 \pm 4$ & $2.3 \pm 0.8$ & $8.1 \pm 2.7$ & $2.3 \pm 0.6$ & $6.0 \pm 1.5$ & $9.9 \pm 4.3$ & $13 \pm 5$\\
  & 20 & $25 \pm 5$ & $2.9 \pm 1.1$ & $11 \pm 3$ & $2.4 \pm 0.7$ & $6.2 \pm 1.8$ & $12 \pm 6$ & $17 \pm 7$\\
  & 40 & $41 \pm 8$ & $4.8 \pm 1.8$ & $17 \pm 6$ & $3.7 \pm 1.1$ & $9.5 \pm 2.7$ & $13 \pm 6$ & $18 \pm 8$\\
 B4 & 10 & $25 \pm 5$ & $8.8 \pm 3.0$ & $6.9 \pm 2.6$ & $4.3 \pm 1.1$ & $3.7 \pm 1.0$ & $20 \pm 9$ & $19 \pm 9$\\
  & 20 & $32 \pm 6$ & $11 \pm 4$ & $8.9 \pm 3.4$ & $4.8 \pm 1.4$ & $4.4 \pm 1.3$ & $23 \pm 11$ & $20 \pm 10$\\
  & 40 & $51 \pm 10$ & $18 \pm 6$ & $15 \pm 6$ & $7.6 \pm 2.3$ & $7.0 \pm 2.1$ & $24 \pm 11$ & $21 \pm 10$\\
\cutinhead{central component (C)}
 CMM3 & 10 & $14 \pm 3$ & $2.1 \pm 0.9$ & $4.4 \pm 1.7$ & $1.9 \pm 0.5$ & $3.6 \pm 1.0$ & $11 \pm 6$ & $12 \pm 6$\\
  & 20 & $18 \pm 4$ & $2.7 \pm 1.1$ & $5.7 \pm 2.2$ & $2.2 \pm 0.7$ & $4.2 \pm 1.2$ & $12 \pm 6$ & $14 \pm 7$\\
  & 40 & $29 \pm 6$ & $4.5 \pm 1.9$ & $9.4 \pm 3.6$ & $3.5 \pm 1.1$ & $6.6 \pm 2.0$ & $13 \pm 7$ & $14 \pm 7$\\   
 CMM13 & 10 & $10 \pm 2$ & $1.1 \pm 0.5$ & $2.4 \pm 0.9$ & $1.9 \pm 0.5$ & $4.9 \pm 1.1$ & $5.8 \pm 3.2$ & $4.9 \pm 2.1$\\
  & 20 & $13 \pm 3$ & $1.4 \pm 0.7$ & $3.2 \pm 1.1$ & $2.2 \pm 0.6$ & $4.6 \pm 1.3$ & $6.6 \pm 3.7$ & $6.8 \pm 3.1$\\
  & 40 & $22 \pm 5$ & $2.4 \pm 1.1$ & $5.2 \pm 1.9$ & $3.4 \pm 1.0$ & $7.0 \pm 2.1$ & $6.9 \pm 3.8$ & $7.3 \pm 3.4$\\
 C1 & 10 & $10 \pm 2$ & $0.9 \pm 0.4$ & $2.8 \pm 1.1$ & $2.0 \pm 0.5$ & $5.6 \pm 1.3$ & $4.8 \pm 2.2$ & $4.9 \pm 2.2$\\
  & 20 & $13 \pm 3$ & $1.2 \pm 0.5$ & $3.6 \pm 1.4$ & $1.9 \pm 0.5$ & $5.2 \pm 1.5$ & $6.5 \pm 3.1$ & $6.8 \pm 3.2$\\
  & 40 & $21 \pm 5$ & $2.0 \pm 0.8$ & $5.9 \pm 2.3$ & $2.9 \pm 0.9$ & $8.0 \pm 2.4$ & $6.9 \pm 3.4$ & $7.3 \pm 3.6$\\
 C2 & 10 & $12 \pm 3$ & $1.9 \pm 0.9$ & $3.2 \pm 1.3$ & $3.9 \pm 0.9$ & $7.9 \pm 1.6$ & $4.8 \pm 2.6$ & $4.1 \pm 1.8$\\
  & 20 & $15 \pm 3$ & $2.4 \pm 1.2$ & $4.2 \pm 1.7$ & $3.9 \pm 1.1$ & $6.4 \pm 1.8$ & $6.3 \pm 3.5$ & $6.5 \pm 3.1$\\
  & 40 & $25 \pm 5$ & $3.9 \pm 1.9$ & $6.8 \pm 2.7$ & $5.9 \pm 1.7$ & $9.5 \pm 2.8$ & $6.7 \pm 3.8$ & $7.1 \pm 3.5$\\
 C3 & 10 & $10 \pm 2$ & $1.2 \pm 0.6$ & $3.1 \pm 1.2$ & $3.2 \pm 0.8$ & $19 \pm 2$ & $3.6 \pm 2.1$ & $1.6 \pm 0.7$\\
  & 20 & $14 \pm 3$ & $1.5 \pm 0.8$ & $4.0 \pm 1.6$ & $3.3 \pm 0.9$ & $8.3 \pm 2.2$ & $4.6 \pm 2.9$ & $4.9 \pm 2.3$\\
  & 40 & $22 \pm 5$ & $2.5 \pm 1.4$ & $6.6 \pm 2.7$ & $5.0 \pm 1.5$ & $12 \pm 3$ & $4.9 \pm 3.1$ & $5.5 \pm 2.7$\\
\enddata
\tablenotetext{a}{Total column density of SiO, obtained by integrating the emission over the entire velocity range}
\tablenotetext{b}{---: the observed line flux is under-reproduced at the indicated value of $T_\mathrm{ex}$}
\end{deluxetable}

\subsection{Molecular abundances}\label{abus}

Finally, we estimated the SiO abundance at each selected position. For the narrow component, we adopted a fixed HCO$^+$ abundance across the region of $5 \times 10^{-9}$ with respect to H$_2$. While this may introduce some uncertainty in the abundance derivation, several studies show that the abundance of HCO$^{+}$ does not vary significantly from this value bewteen outflows and dense cores (see \citealp{alvaro13} and references therein). Since we observed the $^{13}$C isotopologue of HCO$^+$, we adopted $^{12}$C/$^{13}$C\,=\,60, as in Watanabe et al. \cite{wata15}. With these assumptions, the derived SiO abundances are on the order of a few times 10$^{-11}$ for the narrow component. The values are reported in column\,6 of Table\,\ref{tcoln}.

Estimates of the CH$_3$OH abundance are also provided in column 7 of Table\,\ref{tcoln} for the narrow component. The methanol abundance associated with the high-velocity gas cannot be reliably determined due to line blending with two other transitions of the same species, as mentioned earlier.

Since H$^{13}$CO$^+$ displays no detectable extended line wings, we cannot use this tracer to determine the SiO abundance of the high-velocity gas for the broad and central components. Instead, we have used the column densities obtained for the extended wings of the CO lines, by adopting a CO abundance of $10^{-4}$ with respect to H$_2$. The SiO abundances derived in this manner are listed in columns\,8 and 9 of Table\,\ref{tcolb}, for the blue- and red-shifted line wings, respectively. They are on the order of $10^{-9} - 10^{-8}$, with the central component having smaller abundances than the broad component by a factor $2-3$. For cosistency, we also derived a lower limit to the SiO abundance using a 2$\sigma$ upper limit on the H$^{13}$CO$^+$ column density at high velocities. The resulting values are on the order of $10^{-9}$ or smaller, and therefore consistent with the abundances obtained using the CO column densities.

Table\,\ref{tmean} summarizes, for each component, the mean column densities and abundances of the sample of selected positions in Fig.\,\ref{fratio}, along with their standard deviations. For the broad and central components, two SiO column density estimates are provided, as explained in Sect.\,\ref{high}: one that considers only the high-velocity gas, B and C, and one that considers the total amount of gas including ambient and high velocities, B (tot) and C (tot). For the undetected line wings of H$^{13}$CO$^+$, a 2$\sigma$ upper limit is provided.

Several points can be extracted from this table. First, while the narrow component shows little dispersion in the SiO and H$^{13}$CO$^+$ column densities, resulting in a low dispersion of the mean SiO abundance, the broad component has a more scattered range of SiO and CO column densities, which results in a large dispersion of the mean SiO abundance. The central component has a low scatter in the SiO column density values, but a high one for CO, which yields a large dispersion for the mean SiO abundance. Second, the abundances of SiO and CH$_3$OH vary by less than a factor 2 from $T_\mathrm{ex} = 10$\,K to $T_\mathrm{ex} = 40$\,K. The column densities differ by a factor lower than 2 for all the molecules except CH$_3$OH, for which the difference rises up to a factor 4. Finally and most importantly, the abundance of SiO in the narrow component is more than two orders of magnitude lower than that in the central and broad components. In addition, the central component appears less abundant in SiO than the broad component by an average factor of 2.

\begin{deluxetable}{lcccccc}
\tabletypesize{\footnotesize}
\tablewidth{0pt}
\tablecolumns{7}
\tablecaption{Average molecular column densities and abundances for each component\tablenotemark{a}\label{tmean}}
\tablehead{
\colhead{Comp.\tablenotemark{b}}  & \colhead{$\bar{N}_\mathrm{SiO}$} & \colhead{$\bar{N}_\mathrm{CH3OH}$} & \colhead{$\bar{N}_\mathrm{H13CO+}$} & \colhead{$\bar{N}_\mathrm{CO}$} & \colhead{$\bar{\chi}_\mathrm{SiO}$\tablenotemark{c}} & \colhead{$\bar{\chi}_\mathrm{CH3OH}$\tablenotemark{c}}\\
\colhead{}  & \colhead{($10^{12}$\,cm$^{-2}$)}  & \colhead{($10^{14}$\,cm$^{-2}$)} & \colhead{($10^{12}$\,cm$^{-2}$)}  & \colhead{($10^{16}$\,cm$^{-2}$)} & \colhead{($10^{-11}$)}  & \colhead{($10^{-9}$)}}
\startdata
\cutinhead{$T_\mathrm{ex} = 10$\,K}
N & $1.8 \pm 0.4$ & $3.6 \pm 0.6$ & $3.4 \pm 0.5$ & --- & $4.5 \pm 0.5$ & $9.0 \pm 1.1$\\
B & $5.8 \pm 3.1$ & --- & $< 6.8$ & $5.8 \pm 3.6$ & $1300 \pm 1200$ & ---\\
B (tot) & $20 \pm 5$ & --- & $6.6 \pm 2.4$ & --- & --- & ---\\
C & $2.3 \pm 1.0$ & --- & $< 4.3$ & $5.4 \pm 5.0$ & $580 \pm 510$ & ---\\
C (tot) & $11 \pm 1$ & --- & $9.8 \pm 0.8$ & --- & --- & ---\\
\cutinhead{$T_\mathrm{ex} = 20$\,K}
N & $2.4 \pm 0.5$ & $6.4 \pm 0.9$ & $4.7 \pm 0.7$ & --- & $4.3 \pm 0.4$ & $12 \pm 1$\\
B & $7.5 \pm 4.0$ & --- & $< 11.7$ & $5.7 \pm 2.4$ & $1400 \pm 1400$ & ---\\
B (tot) & $25 \pm 7$ & --- & $8.7 \pm 3.0$ & --- & --- & ---\\
C & $3.0 \pm 1.4$ & --- & $< 6.1$ & $4.2 \pm 1.9$ & $750 \pm 620$ & ---\\
C (tot) & $15 \pm 2$ & --- & $13 \pm 1$ & --- & --- & ---\\
\cutinhead{$T_\mathrm{ex} = 40$\,K}
N & $3.9 \pm 0.8$ & $15 \pm 2$ & $7.9 \pm 1.1$ & --- & $4.1 \pm 0.4$ & $16 \pm 2$\\
B & $12 \pm 6$ & --- & $< 17$ & $8.6 \pm 3.3$ & $1500 \pm 1400$ & ---\\
B (tot) & $41 \pm 11$ & --- & $14 \pm 5$ & --- & --- & ---\\
C & $4.9 \pm 2.2$ & --- & $< 10$ & $6.4 \pm 2.8$ & $800 \pm 650$ & ---\\
C (tot) & $24 \pm 3$ & --- & $21 \pm 2$ & --- & --- & ---\\
\enddata
\tablenotetext{a}{The values given are the mean of all the selected positions for each component, and its standard deviation. For CH$_3$OH and H$^{13}$CO$^+$, only the narrow component is considered, while for CO only the broad and central components are taken into account (see text).}
\tablenotetext{b}{Column densities for B (tot) and C (tot) refer to those integrated across the whole velocity range of the emission line. Column densities for B and C refer to those of the blue- and red-shifted high-velocity emission (see text).}
\tablenotetext{c}{Abundance with respect to H$_2$}
\end{deluxetable}

\section{Discussion}\label{discussion}

In the present section, we discuss the narrow, broad, and central components on the basis of our results and those reported in the literature.

\begin{figure}
\epsscale{0.45}
%\epsscale{0.9}
\plotone{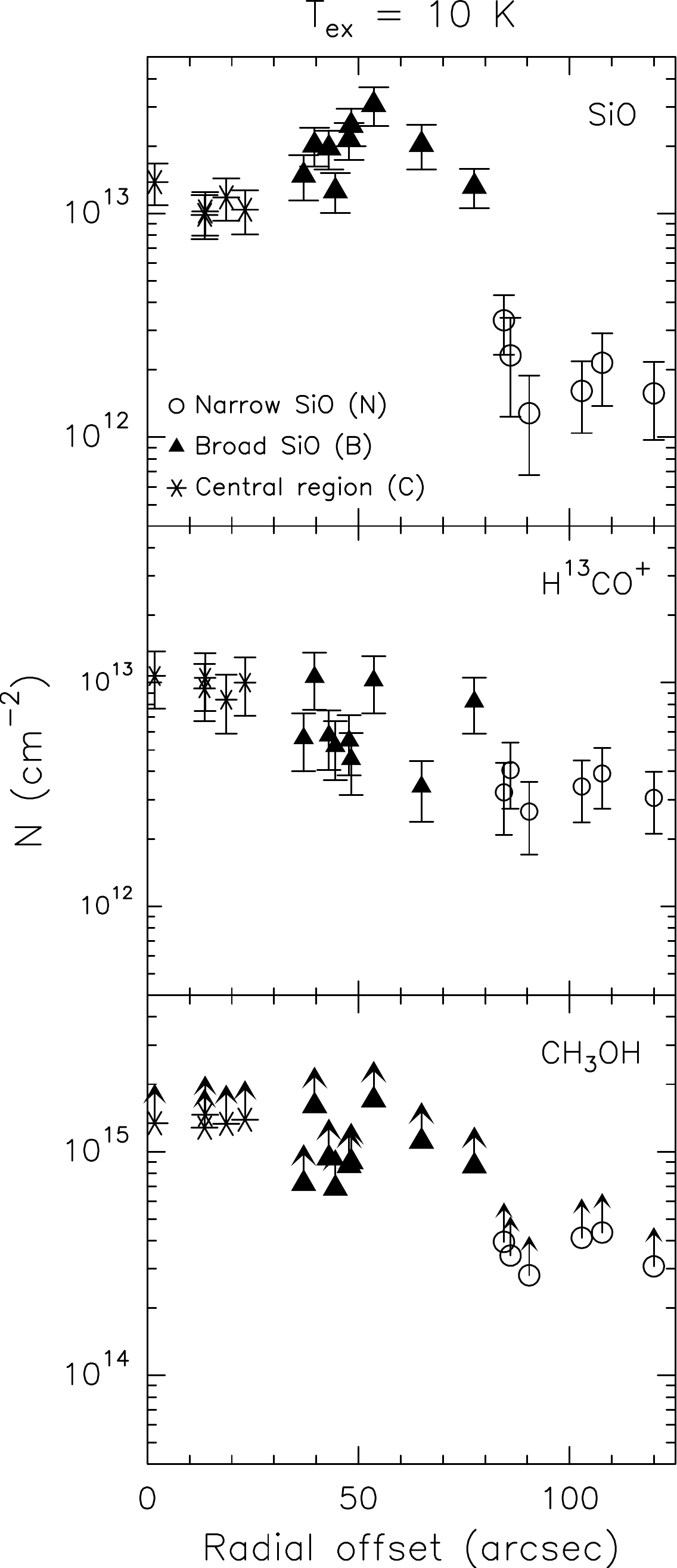}
\caption{Total column density of SiO (\textit{top}), H$^{13}$CO$^+$ (\textit{middle}), and CH$_3$OH (\textit{bottom}), as a function of radial offset with respect to the reference position, for an excitation temperature of 10\,K. Note that, for CH$_3$OH, the values are lower limits, particularly for the broad and central components, since only the central velocities were considered in the column density derivation. The reference (0$''$,0$''$) position corresponds to R.A.(J2000)\,=\,06$^\mathrm{h}$41$^\mathrm{m}$12.3$^\mathrm{s}$, Dec.(J2000)\,=\,09$^\circ$29$'$11.90$''$.
\label{fradial}}
\end{figure}

\subsection{Central SiO depletion}\label{depletion}

Figure\,\ref{fradial} shows the total column densities of SiO, H$^{13}$CO$^+$, and CH$_3$OH, derived assuming $T_\mathrm{ex} = 10$\,K for each selected position in Table\,\ref{tflux}, and plotted as a function of radial distance from CMM3, i.e. roughly the center of the clump. The plots for $T_\mathrm{ex} = 20$ and 40\,K are very similar qualitatively. The SiO plot clearly highlights the three distinct components that we identified in Sect.\,\ref{3comp}. If we concentrate on the shortest radial distances, it is evident that the column density is almost a factor 3 lower within a radius of $25''$ (0.09\,pc) from the center than at larger distances up to about 60$''$ ($0.22$\,pc), where the strongest SiO emission can be seen in Fig.\,\ref{fsptmap1}\footnote{Notice that there is a slight enhancement at the position of CMM3, but this is likely due to the presence of a compact outflow driven by this source (Saruwatari et al.~\citealp{saru11}).}. This decrease in the amount of gaseous SiO towards the center of the protocluster contrasts with the column density profile of H$^{13}$CO$^+$, which peaks at the center and gradually decreases with increasing radial distance.

Therefore, it appears that the central area of NGC\,2264-C has a relative depletion of SiO despite it showing signs of protostellar-outflow activity, as acknowledged from (i) the high-velocity wings present in the SiO line profiles and (ii) the CO outflow maps reported by Maury et al.~\cite{maury09}. Depletion of gaseous SiO may occur through freeze-out back onto dust grains, which has a reported timescale of $\sim 10^4$\,years \cite{bergin98}. It may also be destroyed via a neutral-neutral reaction with OH, yielding SiO$_2$, which has no activation barrier. This reaction is slow, with a rate constant of $2 \times 10^{-12}$\,cm$^{3}$\,s$^{-1}$ (KIDA database\footnote{http://kida.obs.u-bordeaux1.fr/}, \citealp{wakelam12}), but it is expected to be effective in outflow regions due to the enhanced abundances of OH in shocks (see e.g. \citealp{cc98}, \citealp{javi06}, where OH abundances of $\sim 10^{-6}$ are measured). The reported timescale of this reaction is also on the order of 10$^4$\,years \cite{pdf97}. However, both in the case of SiO depletion onto dust grains and of SiO gas-phase destruction, the quoted time scales assume a gas volume density of $10^5$\,cm$^{-3}$. The density around the center of NGC\,2264-C is calculated to be higher than 10$^6$\,cm$^{-3}$ and as high as 10$^8$\,cm$^{-3}$ (see Fig.\,9 in Peretto et al.~\citealp{peretto06}). In these high density conditions, the depletion of gaseous SiO is expected to be much faster, as fast as a few hundred years. We may thus be seeing the result of partial destruction/depletion of SiO in this central region.

An analogous behavior has been reported in e.g. the low-mass protostar NGC\,1333 IRAS\,2A, where SiO emission becomes weaker very close to the driving source due to the enhanced density \cite{jes04}. However, the radial distances involved in that case are on the order of $10^3$\,AU and they concern an individual protostar, while in our case we measure a radial distance two orders of magnitude larger: it is a \textit{protocluster} rather than a \textit{protostar} effect. 

Methanol (CH$_3$OH), on the other hand, does not show evidence of central depletion in Fig.\,\ref{fradial}. However, it should be kept in mind that, for this molecule, we have only integrated the emission at the central line velocities to obtain its column density. The values in this plot should be thus taken as lower limits and are subject to larger errors than the other two molecules. The column densities derived for the narrow (N) component are most likely closer to the actual values than those of the central (C) and broad (B) components, since the line widths are narrower and thus less line flux has been missed in the calculation. Nevertheless, this plot, as well as the CH$_3$OH map in Fig.\,\ref{fsptmap1}, indicates that this molecule's emission is relatively stronger at the center than SiO. Therefore, CH$_3$OH does not suffer the same kind of partial depletion as SiO in the inner region of the protocluster. This may be due to the higher degree of volatility of CH$_3$OH and the high temperatures close to the presumed hot core within CMM3 (Watanabe et al.~\citealp{wata15}), which favor the direct vaporisation of methanol from grain mantles through heating. Indeed, methanol will be significantly sublimated from the mantles at temperatures above 85\,K \cite{bb07}. CH$_3$OH may also be released into the gas as a result of protostellar shocks, but depletion back onto dust grains would not occur as rapidly as for SiO, especially if temperatures are not very low (e.g. \citealp{yana14}).

\subsection{Narrow SiO emission: A tracer of a previous star formation episode?}\label{old}

The top panel in Fig.\,\ref{fradial} shows clearly how, beyond a radial offset of 80$''$ (i.e. $\sim 0.3$\,pc) from the center, the total SiO column density decreases drastically by one order of magnitude. The change is even more pronounced in terms of abundance, where the difference is of more than two orders of magnitude with respect to the broad component (see Table\,\ref{tmean}). As discussed above, the SiO lines are also much narrower here, with average Full Widths at Half Maximum ($FWHM$) of 2\,km\,s$^{-1}$ and no detectable high-velocity wings. This kind of SiO line profile has been reported in several other star-forming regions (e.g. Lefloch et al.~\citealp{lf98}, \citealp{hatchell01}, Jim\'enez-Serra et al.~\citealp{js10}, Duarte-Cabral et al.~\citealp{dc14}). As in the present work, all these studies report an SiO abundance with respect to H$_2$ around $10^{-11} - 10^{-10}$ for such a low-velocity component, i.e. much lower than the abundances typically associated with fast outflow shocks ($10^{-9} - 10^{-7}$; e.g. \citealp{bachiller97}, \citealp{alvaro13}, this work). The abundances associated with this narrow SiO emission are, however, higher than those found in dark clouds ($\lesssim 10^{-12}$; \citealp{ziurys89}, \citealp{rt07}). Therefore, something must be causing this slight enhancement of gaseous SiO.

A possible explanation for this low-velocity SiO component, first proposed by Lefloch et al.~\cite{lf98} and Codella et al.~\cite{clau99}, is that this emission traces old protostellar outflow shocks, which have decelerated and become significantly depleted in SiO over time. Codella et al. estimated the deceleration timescale to be on the order of 10$^4$\,years, the same time scale in which SiO would have been largely depleted back onto dust grains and/or destroyed in the gas phase via reactions with OH. This scenario is consistent with both the narrow SiO line profile and the relatively low SiO abundance.

To evaluate whether this could be the case in NGC\,2264-C, we consider both the spatial distribution of the narrow SiO emission in Figs.\,\ref{fsptmap1} and \ref{fratio}, and the dynamical age of the molecular outflows identified by Maury et al.~\cite{maury09}. The narrow SiO component is distributed mostly near the northern and eastern edges of the protocluster, sufficiently close to the tips of several CO outflows identified by Maury et al. These CO outflows have reported dynamical ages of $\sim 10^4$\,years. Therefore, it is possible that the narrow SiO emission is tracing older shocks from such molecular outflows, which have decelerated to ambient velocities. At the radial distance where the low-velocity component is located, the average gas density is on the order of $10^5$\,cm$^{-3}$ (Peretto et al.~\citealp{peretto06}), thus consistent also with a depletion timescale of about 10$^4$\,years \cite{bergin98}. However, as can be seen in Fig.\,\ref{fradial}, the radial change in column density is rather sudden from the broad to the narrow component positions, and the abundance decrease is even steeper. One would expect a more gradual gradient if the low-velocity SiO emission were due to the ageing of the molecular outflow as it travels further away from its driving source.

An alternative explanation for such a drastic spatial change in the SiO abundance is the following. NGC\,2264-C is known to have experienced more than one episode of star formation. It might be possible that the narrow SiO emission seen in this region has its origin in the star formation episode that preceded the one that is currently taking place. If this is so, the astrochemical remnants of that previous episode would have been replaced by those of the current protostellar activity everywhere but in the most external areas of the protocluster, where contamination from current outflows is less important. It is not obvious to unambiguously identify such a stellar population, as it may be deeply embedded in the gas and thus undetectable/unresolvable at the angular resolution of our observations. Furthermore, the known young stars detected in infrared (IR) studies \cite{lada93}, may be the result of an even earlier star formation episode (their ages are 10$^5 - 10^6$\,years), and a K-band source is also known to reside close to CMM3 (\citealp{schreyer97}, \citealp{schreyer03}). These evolved sources are thus unlikely to be responsible for the narrow SiO component detected. Perhaps the previous star-formation episode is related to the B2-type Zero-Age Main Sequence (ZAMS) star, IRS1, to the north-west of NGC\,2264-C (\citealp{allen72}, \citealp{thompson98}). This high-mass star opens up a cavity that is 0.1 pc in diameter, and it is surrounded by a shell of molecular gas that contains the cores CMM5, CMM10, CMM11, and CMM12, which are believed to be the result of star formation triggered by IRS1. While this is merely a speculation, IRS1 might be part of the stellar population that formed before the younger cores at the center and east of the protocluster. Indeed, a large fraction of the narrow SiO emission is located to the north and west of IRS1.

%Alternatively, the fast abundance change between the two SiO regimes might simply be an artificial effect caused by the selection of the positions where we derived the molecular column densities. \textbf{DO: derivation SiO column density in transition zone.}
%
%Our maps also suggest the presence of a CH$_3$OH narrow component, with an average molecular abundance of about $10^{-8}$. This is orders of magnitude lower than the abundances typically associated with protostellar shocks (e.g. \citealp{bachiller95}, \citealp{jes04}). Sanhueza et al. \cite{sanhueza13} also reported such a low-abundance methanol component, whose emission they attributed to desorption from dust grains as a result of the exothermicity of surface addition reactions. We are unable to conclude whether we are detecting the same kind of emission, due to the severe line-blending suffered by methanol in our observations. Moreover, the CH$_3$OH lines that we detect do not appear as narrow as in Sanhueza et al. \cite{sanhueza13}. Therefore, we will not discuss this component any further. 

\subsection{Other possible origins of the narrow SiO emission}\label{other}

We have discussed above how the narrow SiO emission we detect may be tracing a previous episode of star formation in NGC\,2264-C. In this section we explore alternative explanations. For instance, Jim\'enez-Serra et al.~\cite{js10} proposed that such narrow SiO lines could be produced by low-velocity shocks due to large-scale flow collisions during global cloud collapse. The authors argued that shock velocities of $\sim 12$\,km\,s$^{-1}$ could account for the SiO gas abundances of about 10$^{-11}$ observed in the infrared-dark cloud (IRDC) G35.39--00.33. This scenario is favoured also by Duarte-Cabral et al.~\cite{dc14}, who performed interferometric observations of SiO($J = 2-1$) in the high-mass star-forming complex Cygnus\,X. They proposed an evolutionary picture in which the earliest stages of protocluster formation are characterised by extended low-velocity SiO emission that traces shocks from the large-scale collapse of material onto them. At later stages, when star formation becomes more active and single massive protostars are formed, the SiO luminosity is largely dominated by powerful outflows, and the weaker narrow component shows the last remnants of the initial collapse. If this is the case for NGC\,2264-C, we could be witnessing the later evolutionary stages mentioned by Duarte-Cabral et al., where the multiple outflows present dominate the SiO emission, and the last hints of large-scale collapse can only be seen at the outskirts of the region in the form of narrow and weak SiO lines.

Peretto et al.~\cite{peretto06} found that NGC\,2264-C is indeed undergoing global infall along the axis connecting CMM2, CMM3, and CMM4. However, they measure an infall velocity of 1.3\,km\,s$^{-1}$, rather small to produce the observed amounts of narrow SiO emission according to Jim\'enez-Serra et al.~\cite{js10}. Furthermore, most of the narrow SiO emisison is concentrated towards the north of the clump, a direction that is almost perpendicular to that reported for the global infall by Peretto et al.~\cite{peretto06}.

The northern narrow SiO emission might instead be caused by low-velocity shocks due to the interaction of NGC\,2264-C with NGC\,2264-D, which lies about 1\,pc north (see Fig.\,1 in Peretto et al.~\citealp{peretto06}), an option that was considered to be likely in the IRDC G28.23--00.19 by Sanhueza et al. \cite{sanhueza13}. In NGC\,2264-C, however, this may not be the most likely possibility, since the line-of-sight velocity difference between the two clumps is 2\,km\,s$^{-1}$, and there is no clear evidence of cloud-cloud collision between NGC\,2264-C and D in the dynamical analysis reported by Peretto et al.~(\citealp{peretto06}, \citealp{peretto07}).

In summary, the velocities involved in and around NGC\,2264-C seem to be insufficient to produce the observed amounts of SiO in low-velocity shocks due \emph{only} to global collapse or cloud-cloud interaction. However, if there is some excess of silicon in the gas phase, especially if it is in the form of SiO, lower-velocity shocks ($< 10$\,km\,s$^{-1}$) can efficiently produce or maintain the observed amounts of gas-phase SiO (\citealp{n13}, Duarte-Cabral et al.~\citealp{dc14}). As suggested in Sect.\,\ref{old}, SiO is likely enhanced in NGC\,2264-C as a result of earlier high-velocity protostellar shocks. This implies that large-scale low-velocity shocks in this region can only play a role if such SiO remnants from previous protostellar activity are present, which lends support to the scenario that we discussed in Sect.\,\ref{old}.%This possibility should be explored in more detail with larger maps, ideally covering also NGC\,2264-C and at higher-angular resolution than those presented in this work.

Alternatively, as first proposed by Jim\'enez-Serra et al.~\cite{js10}, the low-velocity SiO component could be tracing an unresolved population of low-mass protostars. The beam dilution of the single-dish observations would cause the SiO high-velocity wings to be undetected. We can estimate whether a typical low-mass protostellar outflow might have the line wings of SiO($J = 2-1$) undetected at the distance of NGC\,2264-C and at the angular resolution of our single-dish observations. The high-velocity SiO wings observed in SVS\,13 by Lefloch et al.~\cite{lf98}, for instance, have intensities between 0.1 and 0.5\,K in main-beam temperature units for a beam size of 24$''$ (i.e. 5600\,AU at a distance of 235\,pc; \citealp{hirota08}). Assuming that the low-mass outflow fills such a beam size (as inferred from the maps presented in Lefloch et al.~\citealp{lf98}), and extrapolating it to the distance of NGC\,2264-C, this corresponds to a size of $\sim 7.5''$. At an angular resolution of 20$''$, the same SiO wings would have a beam-diluted intensity of 25--125\,mK, indeed undetectable or just barely detectable at the sensitivity of our observations (see Table\,\ref{tline}). Taking into account that the outflow driven by SVS\,13 is rather powerful and that weaker --and more compact-- outflows may be present, this is therefore a plausible alternative explanation for the narrow SiO emission.

Based on the different arguments exposed, we favor the scenario of old protostellar activity from a previous star formation episode or, alternatively, a population of unresolved low-mass protostars, to explain the narrow SiO emission detected in NGC\,2264-C. Disentangling between these possibilities would clearly benefit from high-angular resolution maps, as well as more SiO transitions to better constrain the properties of the SiO-emitting gas.

%% The displaymath environment will produce the same sort of equation as
%% the equation environment, except that the equation will not be numbered
%% by LaTeX.

\section{Conclusions}\label{conclusions}

In order to study the role of SiO as a tracer of old protostellar shocks in large protoclusters, we have used the Nobeyama\,45\,m telescope to map the high-mass protocluster NGC\,2264-C in SiO($J = 2-1$), as well as H$^{13}$CO$^+$($J = 1-0$), CO($J = 1-0$), and CH$_3$OH(2$_{0} - 1_{0}$~A$^+$), to further constrain the column densities and molecular abundances across the protocluster. Our main results can be summarized as follows.

\begin{enumerate}
\item A complex and tangled network of molecular outflows is present in this region that spans $\sim 0.5$\,pc in size, as evidenced from the spatial distribution of the high-velocity line-wing emission of CO, SiO, and, to a lower extent, CH$_3$OH.
\item We are able to identify three distinct SiO components in NGC\,2264-C: (i) a \textit{broad} component that is spatially offset with respect to the center of the protocluster, characterized by strong SiO emission and high-velocity wings indicative of protostellar outflow activity; (ii) a \textit{central} component within the inner 0.2\,pc of the molecular clump (i.e. around the massive molecular core CMM3), where the gas is densest, characterized by weaker SiO emission than in the broad component, but nevertheless exhibiting high-velocity wings suggestive of outflows; (iii) a \textit{narrow} component, at the periphery of the region, where SiO lines are relatively weak and narrow.
\item While the column densities of H$^{13}$CO$^+$ and CH$_3$OH are enhanced at the center of NGC\,2264-C, SiO appears relatively depleted in this area, with column densities that are a factor 2 to 3 lower than in the more external, broad SiO component. At the high densities associated with the center of the protocluster, SiO destruction/depletion onto dust grains may occur in timescales of only 10$^2 - 10^3$\,years. As a result, SiO appears relatively depleted despite the high outflow activity present. Beyond a radius of $\sim 0.1$\,pc from the center, where the gas density is relatively lower, SiO depletion occurs in much longer timescales ($\sim 10^4$\,years), which explains the stronger SiO emission of the broad component.
\item The SiO abundance with respect to H$_2$ associated with the peripheral narrow component in NGC\,2264-C is $\sim 4 \times 10^{-11}$, i.e. more than two orders of magnitude lower than those associated with the broad and central components ($\sim 1.4 \times 10^{-8}$ and $\sim 4 \times 10^{-9}$, respectively). Such a low abundance of SiO is nevertheless higher than that associated with dark clouds ($\lesssim 10^{-12}$).
\item We propose that the narrow SiO component may be tracing the remnants of the star formation event that preceded the current one in NGC\,2264-C. If this is the case, the narrow SiO line emission would be providing astrochemical evidence that more than one star-formation episode occurred in this region. Alternatively, an unresolved population of low-mass protostars could be responsible for the narrow SiO emission, where the non-detection of high-velocity wings would be due to beam dilution effects of the single-dish observations. Distinguishing between these two possibilities requires high-angular resolution mapping with interferometers.
\item Based on the low large-scale infall velocities associated with NGC\,2264-C ($\sim 1.4$\,km\,s$^{-1}$), it is unlikely that low-velocity shocks from global collapse are responsible for the narrow SiO emission in this source. Similarly, low-velocity shocks due to cloud-cloud collapse can hardly justify the observed narrow SiO emission, since the projected velocity difference between NGC\,2264-C and NGC\,2264-D is of $\sim 2$\,km\,s$^{-1}$. However, these possibilities should not be completely discarded before obtaining larger maps and, ideally, multiple SiO transitions to better constrain the properties of the SiO-emitting gas.
\end{enumerate}

Following the results of the present study, the role of SiO as a tracer of past star-formation events should be further tested in this and other high-mass protocluster regions.


%% Included in this acknowledgments section are examples of the
%% AASTeX hypertext markup commands. Use \url without the optional [HREF]
%% argument when you want to print the url directly in the text. Otherwise,
%% use either \url or \anchor, with the HREF as the first argument and the
%% text to be printed in the second.

\acknowledgments

The authors are grateful to the NRO staff for their excellent support. The present study is supported by Grant-in-Aid from the Japanese Ministry of Education, Culture, Sports, Science, and Technologies (21224002, 25400223, and 25108005). We thank the anonymous referee for kindly devoting his/her time to review the present manuscript.


%% in parentheses, after the keyword, but they will not be verified.

{\it Facilities:} \facility{NRO}.

%\appendix




\clearpage

%% The following command ends your manuscript. LaTeX will ignore any text


\begin{thebibliography}{}
\bibitem[Allen 1972]{allen72} Allen, D. 1972 ApJ, 172, 55
\bibitem[Anderl et al. 2013]{anderl13} Anderl, S., Guillet, V., Pineau des For\^ets, G., Flower, D. R. 2013, A\&A, 556, A69
\bibitem[Bachiller et al. 1995]{bachiller95} Bachiller, R., Liechti, S., Walmsley, C. M., Colomer, F. 1995, A\&A, 295, 51
\bibitem[Bachiller \& P\'erez-Guti\'errez 1997]{bachiller97} Bachiller, R., P\'erez-Guti\'errez, M. 1997, ApJ, 487, 93
\bibitem[Bergin et al. 1998]{bergin98} Bergin, E. A., Melnick, G. J., Neufeld, D. A. 1998, ApJ, 499, 777
\bibitem[Brown \& Bolina 2007]{bb07} Brown, W. A., Bolina, A. S. 2007, MNRAS, 374, 1006
\bibitem[Ceccarelli et al. 1998]{cc98} Ceccarelli, C., Caux, E., White, G. J. 1998, A\&A, 331, 372
\bibitem[1999]{clau99} Codella, C., Bachiller, R., Reipurth, B. 1999, A\&A, 343, 585
\bibitem[Codella et al. 2007]{clau07} Codella, C., Cabrit, S., Gueth, F. et al. 2007, A\&A, 462, 53
\bibitem[2014]{dc14} Duarte-Cabral, A., Bontemps, S., Motte, F. 2014, A\&A, 570, A1
\bibitem[Goicoechea et al. 2006]{javi06} Goicoechea, J. R., Cernicharo, J., Lerate, M. R. 2006, ApJ, 641, L49
\bibitem[Gueth et al. 1998]{gueth98} Gueth, F., Guilloteau, S., Bachiller, R. 1998, A\&A, 333, 287
\bibitem[Guillet et al. 2009]{guillet09} Guillet, V., Jones, A. P., Pineau Des Forêts, G. 2009, A\&A, 497, 145
\bibitem[Gusdorf et al. 2008]{gusdorf08} Gusdorf, A., Cabrit, S., Flower, D. R., et al. 2008, A\&A, 482, 809
\bibitem[Hatchell et al. 2001]{hatchell01} Hatchell, J., Fuller, G. A., Millar, T. J. 2001, A\&A, 372, 281
\bibitem[Hirota et al. 2008]{hirota08} Hirota, T., Bushimata, T., Choi, Y. K. et al. 2008, PASJ, 60, 37
\bibitem[2010]{js10} Jiménez-Serra, I., Caselli, P., Tan, J. C. et al. 2010, MNRAS, 406, 187
\bibitem[J\o rgensen et al. 2004]{jes04} J\o rgensen, J. K., Hogerheijde, M. R., Blake, G. A. et al. 2004, A\&A, 415, 1021
\bibitem[Kamezaki et al. 2014]{kame14} Kamezaki, T., Imura, K., Omodaka, T. et al. 2014, ApJS, 211, 18
\bibitem[Krumholz et al. 2007]{krum07} Krumholz, M. R.; Klein, R. I.; McKee, C. F. 2007, ApJ, 656, 959
\bibitem[Krumholz \& McKee 2008]{krum08} Krumholz, M. R., McKee, C. F. 2008, Nature, 451, 1082
\bibitem[Krumholz et al. 2014]{krum14} Krumholz, M. R., Bate, M. R., Arce, H. G. et al. 2014, in Protostars and Planets VI, ed. Beuther, H., Klessen, R. S., Dullemond, C. P.  \& Henning, T. (University of Arizona Press, Tucson), 243
\bibitem[Lada \& Lada 2003]{ll03} Lada, C. J., Lada, E. A. 2003, ARA\&A, 41, 57
\bibitem[Lada et al. 1993]{lada93} Lada, C. J., Young, E. T., Greene, T. P. 1993, ApJ, 408, 471
\bibitem[1998]{lf98} Lefloch, B., Castets, A., Cernicharo, J., Loinard, L. 1998, ApJ, 504, 109
\bibitem[2006]{ln06} Li, Z-Y., Nakamura, F. 2006, ApJ, 640, 187
\bibitem[Longmore et al. 2011]{long11} Longmore, S. N., Pillai, T., Keto, E., Zhang, Q., Qiu, K. 2011, ApJ, 726, 97
\bibitem[2009]{ls09} L\'opez-Sepulcre, A., Codella, C., Cesaroni, R., Marcelino, N., Walmsley, C. M. 2009, A\&A, 499, 811
\bibitem[Matzner 2007]{matzner07} Matzner, C. D. 2007, ApJ, 659, 1394
\bibitem[2009]{maury09} Maury, A. J., Andr\'e, Ph., Li, Z.-Y. 2009, A\&A, 499, 175
\bibitem[Nakamura et al. 2011]{nakamura11} Nakamura, F., Kamada, Y., Kamazaki, T. et al. 2011, ApJ, 726, 46
\bibitem[Nakamura \& Li 2007]{nl07} Nakamura, F., Li, Z-Y. 2007, ApJ, 662, 395
\bibitem[Nguyen-Lu'o'ng et al. 2013]{n13} Nguyen-Lu'o'ng, Q.; Motte, F.; Carlhoff, P. et al. 2013, ApJ, 775, 88
\bibitem[2006]{peretto06} Peretto, N., Andr\'e, Ph., Belloche, A. 2006, A\&A, 445, 979
\bibitem[2007]{peretto07} Peretto, N., Hennebelle, P., Andr\'e, Ph. 2007, A\&A, 464, 983
\bibitem[Pineau des For\^ets et al. 1997]{pdf97} Pineau des Forets, G., Flower, D. R., Chieze, J.-P. 1997, in IAUS n.\,182, ed. Reipurth, B. \& Bertout, C. (Kluwer Academic Publishers), 199
\bibitem[Rebull et al. 2002]{rebull02} Rebull, L. M., Makidon, R. B., Strom, S. E. et al. 2002, AJ, 123, 1528
\bibitem[Requena-Torres et al. 2007]{rt07} Requena-Torres, M. A., Marcelino, N., Jim\'enez-Serra, I. et al. 2002, ApJ, 655, L37
\bibitem[S\'anchez-Monge et al. 2013]{alvaro13} S\'anchez-Monge, \'A., L\'opez-Sepulcre, A., Cesaroni, R. et al. 2013, A\&A, 557, 94
\bibitem[2013]{sanhueza13} Sanhueza, P., Jackson, J. M., Foster, J. B. et al. 2013, ApJ, 773, 123
\bibitem[2011]{saru11} Saruwatari, O., Sakai, N., Liu, S-Y. et al. 2011, ApJ, 729, 147
\bibitem[Sawada et al. 2008]{sawada08} Sawada, T., Ikeda, N., Sunada, K. et al. 2008, PASJ, 60, 445
\bibitem[Schreyer et al. 1997]{schreyer97} Schreyer, K., Helmich, F. P., van ishoeck, E. F., Henning, Th. 1997, A\&A, 326, 347
\bibitem[2003]{schreyer03} Schreyer, K., Stecklum, B., Linz, H., Henning, Th. 2003, ApJ, 599, 335
\bibitem[Stanke \& Williams 2007]{sw07} Stanke, T., Williams, J. P.	2007, AJ, 133, 1307
\bibitem[Sung et al. 2008]{sung08} Sung, H., Bessell, M. S., Chun, M-Y., Karimov, R., Ibrahimov, M. 2008, AJ, 135, 441
\bibitem[Sung et al. 1997]{sung97} Sung, H., Bessell, M. S., Lee, S-W. 1997, AJ, 114, 2644
\bibitem[Tafalla et al. 2015]{tafalla15} Tafalla, M., Bachiller, R., Lefloch, B. et al. 2015, A\&A, 573, L2
\bibitem[Thompson et al. 1998]{thompson98} Thompson, R. I., Corbin, M. R., Young, E., Schneider, G. 1998, ApJ, 492, L177
\bibitem[Wakelam et al. 2012]{wakelam12} Wakelam, V., Herbst, E., Loison, J.-C. 2012, ApJS, 199, 21
\bibitem[2015]{wata15} Watanabe, Y., Sakai, N., L\'opez-Sepulcre, A. et al. 2015, ApJ, 809, 162
\bibitem[Yanagida et al. 2014]{yana14} Yanagida, T., Sakai, T., Hirota, T. et al. 2014, ApJ, 794, L10
\bibitem[Ziurys et al. 1989]{ziurys89} Ziurys, L. M., Friberg, P., Irvine, W. M. 1989, ApJ, 343, 201
\end{thebibliography}
\end{document}